\documentclass[11pt,a4paper]{article}
\usepackage{jheppub}
\pdfoutput=1

\usepackage{listings}
\usepackage[section]{placeins}
\usepackage{amsmath}
\usepackage{graphicx}
\usepackage{subfig}
\usepackage{hyperref}

\hypersetup{
   colorlinks=true,       
   linkcolor=blue,        
   citecolor=red,         
   filecolor=magenta,      
   urlcolor=cyan,           
   linktocpage = true,
   }

\newcommand{\stoppingdeltar}{\ensuremath{R}}
\newcommand{\ktstoppingdeltar}{\ensuremath{R_{k_T}}}

\newcommand{\eigenidx}[1]{\textbf{#1}}  
\newcommand{\indicatoridx}[1]{\textbf{#1}}  
\newcommand{\beau}{\ensuremath{b}}
\newcommand{\bbar}{\ensuremath{\bar{b}}}
\newcommand{\bthing}[1]{\ensuremath{b\text{-#1}}}
\newcommand{\genkt}{generalised $k_T$}
\def\antikt{anti-$k_T$}
\newcommand{\itercone}{iterative cone}
\newcommand{\spectral}{spectral}
\def\fastjet{{\tt FASTJET}}

\graphicspath{{./graphics/}}

\allowdisplaybreaks[2]

\title{Spectral clustering for jet physics}

\author[b]{Giorgio Cerro,}
\emailAdd{g.cerro@soton.ac.uk}
\author[a]{Srinandan Dasmahapatra,}
\emailAdd{sd@ecs.soton.ac.uk}
\author[b,c,d]{Henry A. Day-Hall,}
\emailAdd{henry.day-hall@cern.ch}
\author[b]{Billy Ford,}
\emailAdd{b.ford@soton.ac.uk}
\author[b,d]{Stefano Moretti,}
\emailAdd{stefano@phys.soton.ac.uk}
\author[c]{and Claire H. Shepherd-Themistocleous}
\emailAdd{claire.shepherd@stfc.ac.uk}

\affiliation[a]{School of Electronics and Computer Science, University of Southampton, Southampton, SO17 1BJ, United Kingdom}
\affiliation[b]{School of Physics and Astronomy, University of Southampton, Southampton, SO17 1BJ, United Kingdom}
\affiliation[c]{Particle Physics Department, Rutherford Appleton Laboratory, Chilton, Didcot, Oxon OX11 0QX, United Kingdom}
\affiliation[d]{Faculty of Nuclear Sciences and Physical Engineering, Czech Technical University, Prague, 160 00, Czech Republic}

\abstract{
\noindent
We present a new approach to jet definition alternative to clustering methods,
such as the anti-kT scheme, that exploit kinematic data directly.
Instead the new method uses kinematic information to represent the
particles in a multidimensional space, as in spectral clustering.
After confirming its Infra-Red (IR) safety, we compare its performance in analysing
\(gg\to H_{125\,\text{GeV}} \rightarrow H_{40\,\text{GeV}} H_{40\,\text{GeV}} \rightarrow b \bar{b} b \bar{b}\),
\(gg\to H_{500\,\text{GeV}} \rightarrow H_{125\,\text{GeV}} H_{125\,\text{GeV}} \rightarrow b \bar{b} b \bar{b}\)
and
\(gg,q\bar q\to t\bar t\to b\bar b W^+W^-\to b\bar b jj \ell\nu_\ell\) events from 
Monte Carlo (MC) samples, specifically, in reconstructing the relevant final states, to that of the \antikt{} algorithm.
   Finally, we show that the  results for spectral clustering are obtained  without any change in the parameter settings of the algorithm,
   unlike the \antikt{} case, which requires the cone size to be adjusted to the physics process under study.
}


\arxivnumber{2104.01972}

\begin{document}
\maketitle

    \section{Introduction}\label{sec:JetClustering}

To perform jet clustering for hadron collider physics one of three algorithms --  
$k_T$~\cite{Ellis:1993tq}, Cambridge-Aachen (CA)~\cite{Dokshitzer:1997in,Wobisch:1998wt} or \antikt{}~\cite{Cacciari:2008gp, Catani:1993hr, Moretti:1998qx}, 
all of which originated in $e^+e^-$ physics (see  refs.~\cite{Sterman:1977wj,Bethke:1991wk,Catani:1991hj,Moretti:1998qx}) -- is a preferred choice. This is due to several desired properties: they are infrared safe,  are flexible enough to capture many different jet signals with minimal parameter changes and excellent implementations of them are publicly available (see \fastjet{}~\cite{Cacciari:2011ma}).  These algorithms are recursive (or iterative) and agglomerative.
A recursive algorithm is well suited to clustering objects when the number of groups is not known from the outset.
Agglomerative algorithms create jets by grouping objects,  starting from individual particles,
and continuing to combine the groups of particles into larger groups,  until the desired jet size is reached.
Creating jets that are  IR safe can be achieved by ensuring that pairs of  particles emerging from soft and/or collinear emissions combine early in this process.
Once these IR splittings have been recombined they
cannot influence the rest of the clustering process.

Jet definition precedes further algorithmic methods to extract useful
physical quantities. Finding an alternative clustering method that compares favourably to
these popular jet algorithms, and which offers additional features for further analysis, is our goal.
Success in obtaining clusters based on informative transformations of the data
offers the possibility of exploiting such representations.
In this paper, we use Laplacian eigenmaps \cite{Belkin:2003_unfound4} to represent the particles
in an event, a procedure employed in applications such as image segmentation \cite{Shi:1997_unfound595}
and called spectral clustering \cite{Ng:2001_unfound543}.
Spectral clustering has  had success in other physics contexts, such as to identify the motion
of vortices~\cite{Hadjighasem:2016_unfound447} in fluid dynamics. It has also been used to reduce the risk of blackouts in electricity supply, subdividing power grids into `islands'. These are
electromechanically stable regions located by minimising the power flow between them using spectral clustering~\cite{HaoLi:2005_unfound114}.  A hierarchical, agglomerative algorithm for the same was introduced in~\cite{RJSanchezGarcia:2014_unfound420}.  This agglomerative approach is what we show in this paper to be suitable also in the context of jet physics. 

The plan of this paper is as follows. In the next section, we will introduce the fundamentals of the theory of spectral clustering. In the following one, we will describe the details of the specific method that we have applied. The numerical results will then follow. Finally, we will draw our conclusions. 

    \FloatBarrier
    \section{Theory of spectral clustering}\label{sec:spectral_theory}
Gathering soft and/or collinear emissions of particles is the target of jet formation, so this must be decided by localised information.
A representation of observable particles that preserves and accentuates local information
motivates the Laplacian eigenmap~\cite{Belkin:2003_unfound4} and spectral
clustering~\cite{Ng:2001_unfound543}, so as to lead us to believe that these are suitable tools for jet formation.
An excellent description of the theory behind spectral clustering
can be found in~\cite{UlrikevonLuxburg:2007_unfound52} while a short
summary is given in this section.

\begin{figure}[!t]
        \center
        \includegraphics[width=\textwidth]{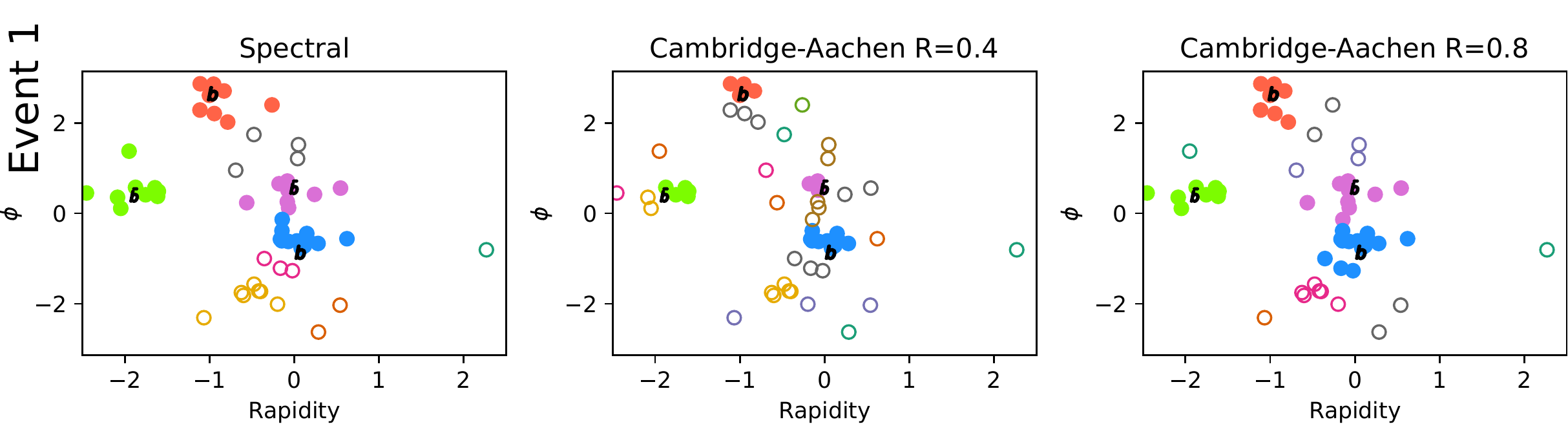}
        \hspace{5mm}
        \includegraphics[width=\textwidth]{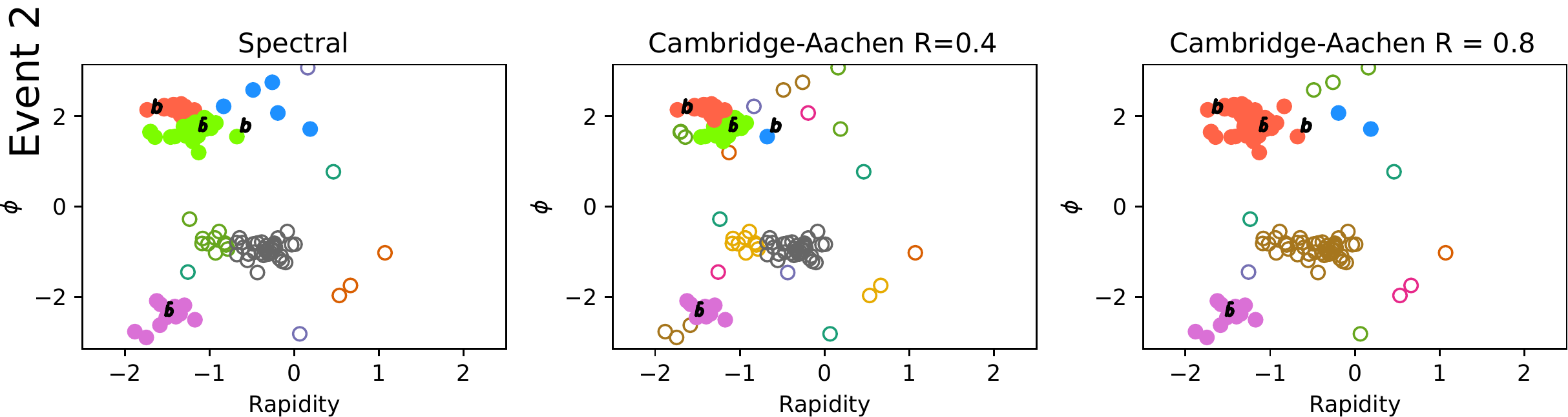}
        \hspace{5mm}
        \includegraphics[width=\textwidth]{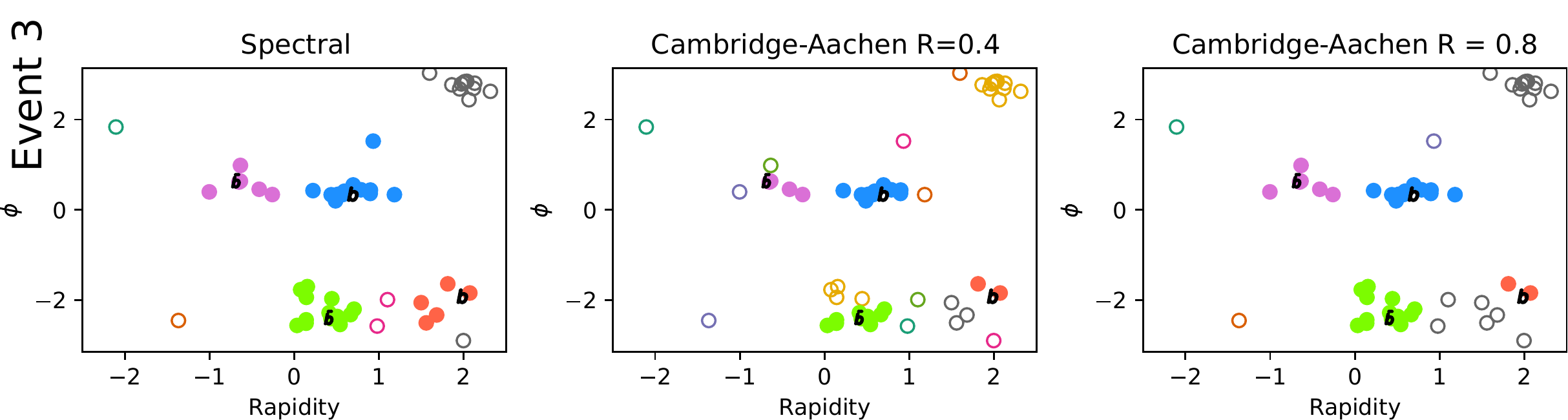}
        \caption{Behaviour of the \spectral{} algorithm is compared to 
                 the well known Cambridge-Aachen algorithm using three events
                 from our dataset.
                 Each row contains an event, each column is a clustering algorithm.
                 Circle colour indicates jet membership, filled circles 
                 indicates a \bthing{quark} jet.
            }\label{fig:real_space}
    \end{figure}

Before looking at the theory behind this algorithm, a quick illustration of
the jets it can produce is shown in figure \ref{fig:real_space}.
This compares the \spectral{} algorithm to the well known CA one using three events.
These events are chosen because they represent different challenging situations for a jet formation algorithm.
Event 1 contains jets that have quite uniform density and blend smoothly with other jets.
Event 2 contains 3 jets in very close proximity. Event 3 contains jets of variable density.
Together their behaviour showcases some of the reasons to be interested in this algorithm.

Spectral clustering is a method whereby a set of points are represented in a new space,
called the embedding space, in which they can be easily clustered.
Coordinates of the
points in the embedding space are expressed in terms of the eigenvectors and eigenvalues
of an associated Laplacian matrix, hence the name.

Input data for spectral clustering must be given as a graph,
which is a set of nodes, in this case representing the particles,
and edges which join nodes together, representing relationships between particles.
The edges may be weighted, that is, a positive number is associated with the edge,
called an affinity.
Affinity represents the degree of belief that the nodes connected by the edge should be in the same group:
for jet clustering this will be a degree of belief that the particles came from the same shower.

The theory behind the
construction of the embedding space is a relaxation
of optimising criteria that would best 
partition nodes into separate disconnected subgraphs, by
splitting nodes into groups.
In a standard (non-physics) procedure we would start from points with coordinates,
which should be split into a predetermined number, \(s\), of clusters.
The points are represented by nodes of a graph.
The edge of the graph joining node (or point) \(i\) and \(j\) has weight \(a_{i, j}\),
which should grow with the probability of \(i\) and \(j\) being in the same group.

To identify groups for the points the graph is split into subgraphs,
\(G_\indicatoridx{k}\), where \(\indicatoridx{k}=1 \dots s\).
These groups should not split up points which are connected by edges with high affinity,
but it should also avoid groups of very uneven size.
Minimising the NCut objective captures this aim, where NCut is defined as
\begin{equation}
    \text{NCut} = \frac{1}{2}\sum_\indicatoridx{k}\frac{W(G_\indicatoridx{k}, \bar{G_\indicatoridx{k}})}{\text{vol}(G_\indicatoridx{k})},
\end{equation}\label{eqn:cost_function}
where \(W(G_\indicatoridx{k}, \bar{G_\indicatoridx{k}})\) is the sum of all the edge weights that must be dropped
to separate the cluster \(G_\indicatoridx{k}\) from the rest of the graph, \(\bar{G_\indicatoridx{k}}\),
so that \( W(G_\indicatoridx{k}, \bar{G_\indicatoridx{k}}) = \sum_{i \in G_\indicatoridx{k}, j \in \bar{G_\indicatoridx{k}}} a_{i, j} \).
In the denominator, \(\text{vol}(G_\indicatoridx{k}) = \sum_{i \in G_\indicatoridx{k}} \sum_{j} a_{i, j}\) is 
the sum of all affinities connecting to a point in \(G_\indicatoridx{k}\).
This denominator is used to penalise the formation of small clusters.

In order to determine which point will go in which \(G_\indicatoridx{k}\), a set of indicator vectors must be found.
Membership of cluster \(G_\indicatoridx{k}\) will be recorded in the indicator vector \(h_\indicatoridx{k}\):
\begin{equation}\label{eqn:indicator}
    h_{\indicatoridx{k}\,i}= 
    \begin{cases}
        1/\sqrt{\text{vol}(G_\indicatoridx{k})}& \text{if point } i \in G_\indicatoridx{k} ,\\
        0             & \text{otherwise}.
    \end{cases}
\end{equation}

To find these indicator vectors the graph is represented by the graph Laplacian, \(L\), a square
matrix with as many rows and columns as there are points.
To construct this Laplacian we define two other matrices:
an off diagonal matrix 
\(A_{i, j} = (1 - \delta_{i, j})a_{i, j}\)
and a diagonal matrix
\(D_{i, j} = \delta_{i, j}\sum_q a_{i, q}\).
Then the symmetric Laplacian can be simply written as
\begin{equation}\label{eqn:symmetric_laplacian}
    L = D^{-\frac{1}{2}} (D - A) D^{-\frac{1}{2}}.
\end{equation}

Considering just one cluster, \(G_\indicatoridx{k}\), when the Laplacian is multiplied by its indicator vector,
the result is the term that NCut seeks to minimise for that cluster,
\begin{equation}
    h_\indicatoridx{k}'Lh_\indicatoridx{k} = \frac{1}{\text{vol}(G_\indicatoridx{k})}\sum_{i \in G_\indicatoridx{k}, j \in G_\indicatoridx{k}} \left(\delta_{i, j}\sum_{l} a_{l, i} - a_{i, j} \right) = \frac{W(G_\indicatoridx{k}, \bar{G_\indicatoridx{k}})}{\text{vol}(G_\indicatoridx{k})}.
\end{equation}
To obtain the sum of all the terms, stack the indicator vectors into a matrix,
\( h'_\indicatoridx{k} L h_\indicatoridx{k} = (H'L H)_{\indicatoridx{k}\indicatoridx{k}}\),
and the NCut aim described earlier becomes the trace
\begin{equation} \text{NCut}(G_\indicatoridx{1},G_\indicatoridx{2}, \dots G_\indicatoridx{n}) \equiv \frac{1}{2} \sum_{k=1}^n \frac{W(G_\indicatoridx{k}, \bar{G_\indicatoridx{k}})}{\text{vol}(G_\indicatoridx{k})} = \text{Tr}(H'LH),\end{equation}
where \(H'H = I\).
This is still a Non-deterministic Polynomial (NP)-hard problem~\cite{Leeuwen:1990_unfound0}.
However, if we relax the requirements made on \(h\) in eq.~(\ref{eqn:indicator}),
allowing the elements of \(h\) to take arbitrary values, then the Rayleigh-Ritz theorem provides a solution.
Trace minimisation in this form is done by finding the eigenvectors of \(L\) with smallest eigenvalues,
\begin{equation}
    \lambda_\text{min} = \min_{\|x\|\ne 0 } \frac{x^H L x}{x^H x},
\end{equation}
where \(x\) is the relaxed indicator vector and an eigenvector of \(L\).
Notice that \(L\) is a real symmetric matrix
and, therefore, all its eigenvalues are real.
Due to the form of the Laplacian, there will be an eigenvector with components all of the same value and its eigenvalue will be \(0\).
This corresponds to the trivial solution of considering all points to be in one group.
The next \(c=s\) eigenvectors of \(L\), sorted by smallest eigenvalue, can be used to allocate points to \(s\) clusters.

These eigenvectors are then used to determine the position of the points in the embedding space.
Each eigenvector has as many elements as there are points to be clustered,
so the coordinates of a point are the corresponding elements of the eigenvectors.

The standard method above is designed to form a fixed number of clusters,
but typically we do not know how many jets should be created in an event.
We will create an alternative algorithm, beginning with the principles of
spectral clustering and adjusting to the needs of the physics being studied.
Using the positions in embedding space, the points can be gathered agglomeratively,
so that we do not need to choose a predetermined number of clusters.

\subsection{Distance in the embedding space}\label{sec:embedding_distance}
When the relaxed spectral clustering algorithm is used to create an embedding space,
points in each group will be distributed in this embedding space.
Each point can be seen as a vector, its direction  indicating the group to which this point should be assigned.
Changes in magnitude of the vectors cause the Euclidean distance between the corresponding points to grow,
however, an angular distance is invariant to changes in magnitude,
 therefore it  is a suitable measure to use.

\subsection{Information in the eigenvalues}\label{sec:eig_norm}
When the clusters in the data are well separated,
the affinities between groups are close to \(0\)
and the eigenvalues will also be closer to \(0\).
So a small eigenvalue means that the corresponding eigenvector
is separating the particles cleanly according to the affinities.
It is possible to make use of this information.

In a traditional application of spectral clustering, the number of clusters desired, \(s\), is predetermined.
The embedding space is created by taking \(c=s\) eigenvectors with smallest eigenvalues, excluding the trivial eigenvector.
The embedding space then has \(c\) dimensions.

When forming jets we do not know from the outset how many clusters to expect in the dataset,
so the number of eigenvectors to keep is not clear.
We cannot set \(c=s\).
While we could choose a fixed, arbitrary number of eigenvectors, this is suboptimal.
A better approach is to take all non-trivial eigenvectors corresponding to eigenvalues
smaller than some limiting number, \(\lambda_\text{limit}\).
For a symmetric Laplacian the eigenvalues are \(0 \leq \lambda_\eigenidx{1} \leq \lambda_\eigenidx{2} \leq \cdots \lambda_\eigenidx{n} \leq 2\),
and \(\lambda_\eigenidx{k}\) is related to the quality of forming \(\eigenidx{k}\)
clusters~\cite{JamesRLee:2014_unfound736}.
Removing eigenvectors with eigenvalues close to \(0\) would result
in discarding useful information, while retaining eigenvectors 
whose eigenvalues are close to \(2\) would increase the noise.
Values of \(0 < \lambda_{\mathrm{limit}} < 1\) are sensible choices
and within this range the choice is not critical.
Then, the number of dimensions in the embedding space will vary,
according to the number of non-trivial eigenvectors with corresponding \(\lambda < \lambda_\text{limit}\).

There is one more manipulation from the information in the eigenvalues.
The dimensions of this embedding space are not of equal importance.
This can be accounted for by dividing the eigenvector by some power, \(\beta\), of the eigenvalue.

Let the eigenvectors for which \(\lambda < \lambda_\text{limit}\) be
\begin{equation}
    \sum_j L_{i, j} x_{\eigenidx{n}\,j} = \lambda_\eigenidx{n} x_{\eigenidx{n}\,i}.
\end{equation}
Then, the coordinates of the \(j^\text{th}\) point in the \(c\) dimensional embedding space
become \(m_j = \left(\lambda_\eigenidx{1}^{-\beta} h_{\eigenidx{1}\,j}, \dots \lambda_\eigenidx{c}^{-\beta} h_{\eigenidx{c}\,j}\right)\).
In effect, the magnitudes of the vectors, \(m_j\), in the \(n^\text{th}\) dimension are compressed by a factor \(\lambda_\eigenidx{n}^\beta\),
so the larger \(\lambda_\eigenidx{n}\) the greater the compression.

\subsection{Stopping conditions}\label{sec:stopping_condintion}

If a recursive algorithm is to be chosen, like the \genkt{} algorithm, a stopping condition is needed.
A stopping condition based on smallest distance between points in the embedding space was attempted 
but this was not found to be stable.
Choosing an acceptable value for all events was not possible.

Distance between the last two points to be joined before the desired jets have been formed
 varies significantly between events, so minimum separation is not a good stopping condition.
The average distance between points before this last joining is more stable because it 
is balanced by two opposing influences.
When points are joined together in a fix number of dimensions the average
distance between points rises.
If this were used in physical space it would be roughly proportional to the number of points remaining.
So, in physical space, if we stopped clustering when the average distance exceeded some cut-off,
we would expect roughly the same number of jets in each event.
However, the embedding space has a variable number of dimensions.
When lots of clustering still remains to be done the lower eigenvalues mean that
the embedding space has more dimensions,
as described in section~\ref{sec:eig_norm}.
When the number of dimensions in the embedding space falls,
the mean distance between points will also fall.

As points combine the mean distance will rise,
but when fewer combinations with higher affinity remain the number of
dimensions in the embedding space falls, counteracting the rise in mean distance.
In short, the mean distance in the embedding space makes a natural cut-off.
The assertions made here are evidenced in appendix~\ref{app:stopping_conditions}.

    \FloatBarrier
    \section{Method}
In this section the methodology is covered in four parts.
Firstly, the algorithm chosen in this work for applying \spectral{} clustering  is given.
Secondly, choices and interpretations for the variable parameters in this algorithm are given.
Thirdly,  the datasets against which this method will be measured are specified.
Fourthly, the procedure for checking functional IR safety is described.

    \begin{figure}[!t]
        \center
        \includegraphics[width=\textwidth]{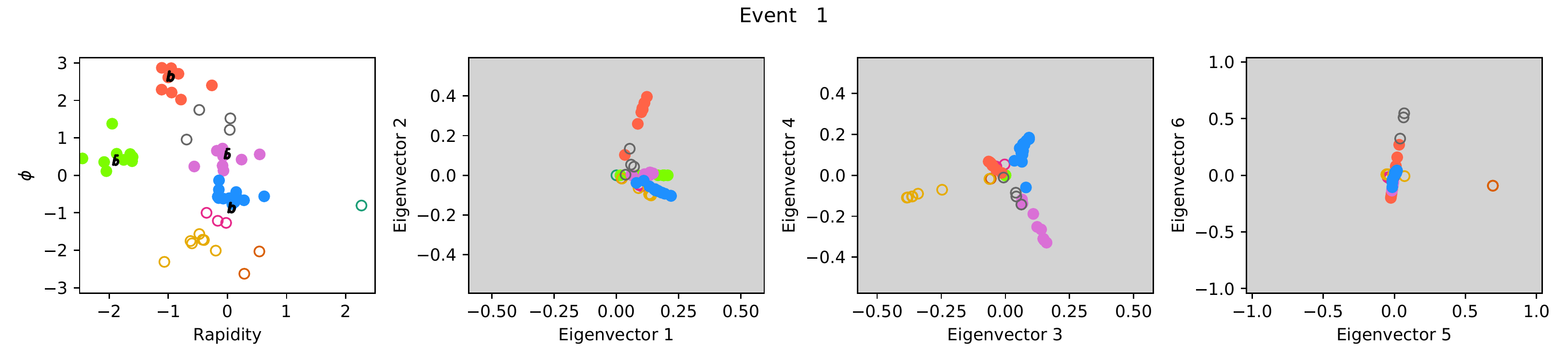}
        \hspace{-5mm}
        \includegraphics[width=\textwidth]{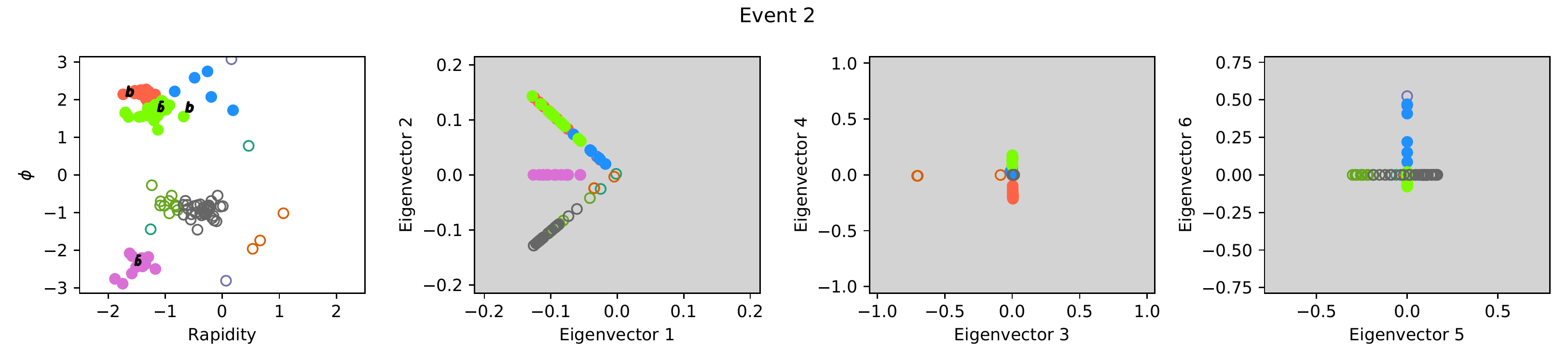}
        \hspace{-5mm}
        \includegraphics[width=\textwidth]{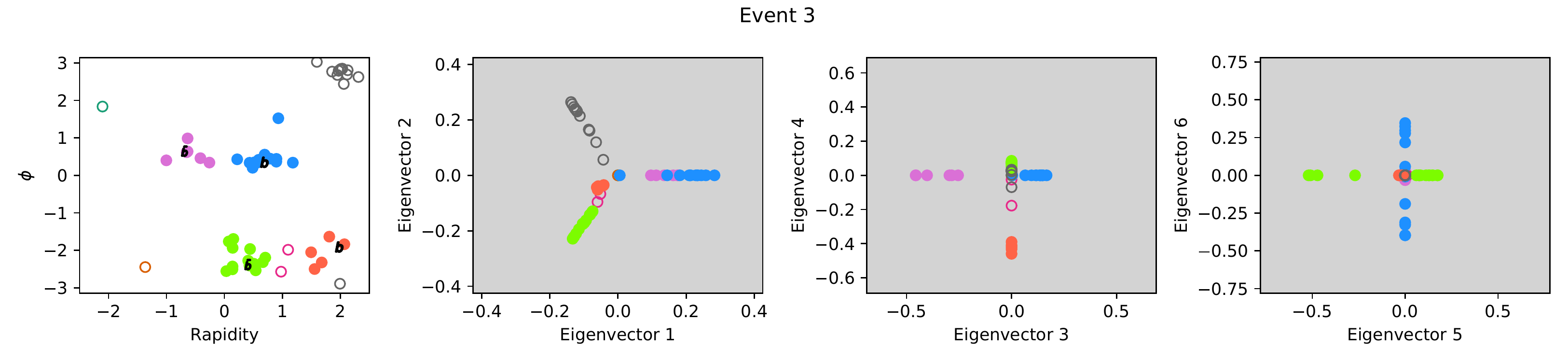}
        \caption{The same three events as in figure~\ref{fig:real_space} 
            are used to show the embedding space created by \spectral{} clustering.
            To the left the white plot shows the particles in the event as points on the unrolled detector barrel.
            The colour of each point indicates the jet it is assigned to, filled circles are
            \bthing{jets}.
            On the right, three grey plots show the first 6 dimensions of the embedding space
            and the location of the points within the embedding space.
            The events show the varying levels of clarity in the embedding space.
            }\label{fig:embedding_space_simple}
    \end{figure}    

Our coordinate system has a \(z\) axis parallel to the beam line.
Directions perpendicular to the \(z\) axis are termed transverse (\(T\)) and
the angle in the transverse plane is labelled \(\phi\).
For clarity, note that the variable pseudorapidity is never used
in the algorithm proposed, all references to rapidity $y$ correspond to
\begin{equation}\label{eqn:rapidity}
    y = \frac{1}{2} \ln\frac{E + p_z}{E - p_z}.
\end{equation}
Rapidity and \(\phi\) form an orthogonal coordinate system.
Distances in the \(\phi\) coordinate must respect its cyclic nature.
The shortest distance between \(\phi_i\) and \(\phi_j\) is denoted as
\begin{equation}\label{eqn:cyclic_phi}
    \delta(\phi_i, \phi_j).
\end{equation}

\subsection{Spectral clustering algorithm}\label{sec:spectralmethodalgo}
    For every simulated event, the following process is used to identify the jets.
    To begin with, relevant cuts are applied to the particles to simulate the detector
    reconstruction capability.
    (These are described in detail in section~\ref{sec:particle_data}.)
    Then all particles are declared pseudojets  and given an index, \(j = 1 \dots n\), with no particular order.
    The algorithm is agglomerative, recursively selecting pairs of pseudojets to merge,
    hence, the first iteration step is labelled \(t=1\).

    When the two pseudojets to be merged, \(i\) and \(j\), have been identified, they are combined
    using the E-scheme.
    The E-scheme forms a new pseudojet by summing the \(4\)-momenta of the two joined pseudojets,
    \(p(t+1)_k = p(t)_i + p(t)_j\).
    The steps used to select two pseudojets to merge proceed as follows.

    \begin{enumerate}
        \item\label{step:start} The pseudojets are used to form the nodes of a graph,
        the edges of which will be weighted by some measure of proximity between the particles called affinity.
        To obtain an affinity, first a distance is obtained.
        Between pseudojets \(i\) and \(j\) this is \(d(t)_{i,j} = \sqrt{(y(t)_i - y(t)_j)^2 + \delta(\phi(t)_i, \phi(t)_j)^2}\),
        where \(y(t)_j\) is the rapidity of pseudojet \(j\) at step \(t\) and \(\phi(t)_j\) is the angle in the transverse plane, likewise for \(i\).
        No \(p_T\) (transverse momentum) dependence is used, unlike in many traditional jet clustering methods.

    \item Calculate a singularity factors for all possible merges.
        This should be \(0\), if the merge could include a soft  particle or a collinear pair,
        or tend to \(1\), otherwise. We use
        \begin{equation}
            s_{i, j}(t) = 1 - \frac{\kappa}{\kappa + \mathrm{min}(p_T(t-1)_i, p_T(t-1)_j)d(t-1)_{i,j}},
        \end{equation}
        where \(\kappa\) is a constant, here chosen to be \(0.0001\).
    \item \label{step:affinity} The affinity must increase as pseudojets become more similar,
        whereas the distance, \(d(t)_{i,j}\), will shrink.
        We chose the affinity \(a(t)_{i,j} = \text{exp}(-d(t)_{i,j}^\alpha/\sigma_v)\),
        where \(\alpha=2\) is the standard Gaussian kernel as
        used in~\cite{Belkin:2003_unfound4}.
            Distances much larger than \(\sigma_v\) are only allowed very small affinities,
            thus less influence over the clustering.

    \item\label{step:KNN} Pseudojets that are far apart have low affinity,
        hence are unlikely to be good candidates for combination.
        Removing these affinities reduces noise.
    A fixed number, \(k_\text{NN}\), of neighbours of each pseudojet is 
    preserved while all other affinities are set to zero.
    Thus, when there are more than \(k_\text{NN}\) pseudojets,
    each pseudojet has at least \(k_\text{NN}\) non-zero affinities with other pseudojets.

    \item\label{step:laplacean} These affinities allow the construction of the normalised
            Laplacian, which is proportional to \(-a(t)_{i, j}\)
            in the \(i\)$^{\rm th}$ row and \(j\)$^{\rm th}$ column.
            For ease of notation, let \(z(t)_j\) be a measure of the size a pseudojet \(j\) contributes to a cluster.
            Before the first merge \(z(1)_j = \sum_k a_{j,k}\).
            Then define three square matrices; \(A(t)_{i, j} = (1 - \delta_{i, j}) a(t)_{i, j}\) which is commonly known as the adjacency matrix,
            \(B(t)_{i, j} = \delta_{i, j} b_i = \delta_{i, j} \sum_k a(t)_{i, k}\) which is commonly known as the degree matrix 
            and \(Z(t)_{i, j} = \delta_{i, j} z(t)_i\) which normalises the Laplacian.
            The Laplacian can now be written as
           \begin{equation}\label{eqn:Laplacian}
               L(t) = Z(t)^{-\frac{1}{2}}(B(t) - A(t))Z(t)^{-\frac{1}{2}}.
           \end{equation}
            After each step this Laplacian shrinks by one row and column.
            If pseudojets \(i\) and \(j\) from step \(t-1\) are merging to form pseudojet \(i\) at step \(t\), we have
            \begin{equation}\label{eqn:newweight}
                z(t)_i = s_{i,j}(t-1)(z(t-1)_i + z(t-1)_j) + (1 - s_{i,j}(t-1))b(t)_i.
            \end{equation}
            All other pseudojets take the size \(z(t)_q = s(t)z(t-1)_q + (1 - s(t))b_q\).
            This is designed so that the size of a pseudojet grows cumulatively,
            but it is reset if soft or collinear particles merge into it.

    \item \label{step:eigenvectors} The eigenvectors of \(L(t)\) (\(\eigenidx{q}\) being the eigenvalue index)
            \begin{equation}
                L(t) h(t)_\eigenidx{q} = \lambda(t)_\eigenidx{q} h(t)_\eigenidx{q},  \; \eigenidx{q}=1, \ldots c
            \end{equation}
            	are used to create the embedding of the pseudojets.
            The eigenvector corresponding to the smallest eigenvalue represents the trivial solution,
            which would place all points in the same cluster (see section~\ref{sec:spectral_theory}).
            All non-trivial eigenvectors, corresponding
            to eigenvalues less than an eigenvalue limit, \(\lambda(t)_\eigenidx{c} < \lambda_\text{limit} < \lambda(t)_\eigenidx{c+1}\),
            are retained (see section~\ref{sec:eig_norm}).
            If no eigenvectors are retained by this, the clustering ends here.

        \item \label{step:compression} An eigenvector is divided by the corresponding eigenvalue raised to \(\beta\).
            To prevent zero division errors, the smallest eigenvalues are clipped to \(0.001\),
            such that \(\lambda'_\eigenidx{q} = \min(\lambda_\eigenidx{q}, 0.001)\).
            This acts to compress the dimensions that hold less information, again, see section~\ref{sec:eig_norm}.
            The embedding space can now be formed.
            The eigenvectors have as many elements as there are pseudojets and the coordinates of
            the \(j^\text{th}\) pseudojet at step \(t\)
            are defined to be
            \(m(t)_j = \left(\lambda'_\eigenidx{1}(t)^{-\beta} h_\eigenidx{1}(t)_j, \dots \lambda'_\eigenidx{c}(t)^{-\beta} h_\eigenidx{c}(t)_j\right)\).

        \item  A measure of distance between all pseudojets in the embedding space is calculated.
            In the embedding space angular distances are most appropriate (see section~\ref{sec:embedding_distance}):
            \begin{equation}
                d'(t)_{i, j} = s(t)_{i,j}\arccos\left(\frac{m(t)_i\cdot m(t)_j}{\|m(t)_i\| \|m(t)_j\|}\right),
            \end{equation}
            where \(\|m\|\) is the (Euclidean) length of \(m\).
            The factor of \(s(t)_{i,j}\) ensures that all soft and/or collinear particles are merged early in the clustering.
            This is important because such a merge will reset the size of the pseudojet in step \ref{step:laplacean}.

        \item\label{step:stoppingcondition}

            A stopping condition, based on the parameter \stoppingdeltar{}, is now checked.
            Provided the mean of the square roots of the distances \(d'(t)_{i, j}\) is less than 
            the value of \stoppingdeltar{}, that is,
            \begin{equation}
                \frac{2}{c(c-1)}\sum_{i\ne j} \sqrt{d'(t)_{i, j}} < \stoppingdeltar{},
            \end{equation}
            then the two pseudojets that have the smallest embedding distance are combined.
            (Reasons for this stopping condition are given in section~\ref{sec:stopping_condintion}.)
        
     \end{enumerate}
    When the mean of the distances in the embedding space rises above \stoppingdeltar{},
    then all remaining pseudojets are promoted to jets.
    Jets with less than 2 tracks are removed and their contents considered noise.
    Further cuts may then be applied as described in section~\ref{sec:particle_data}.

    These steps will form a variable number of jets from a variable number of particles.
    An example of the constructed first embedding space is shown in figure~\ref{fig:embedding_space_simple}.
    This illustrates how the embedding space highlights the clusters.

\subsection{Tunable parameters}\label{sec:spectralmethodparam}
Unlike most deep learning methods currently used in particle physics, spectral clustering does not have large arrays of learnt parameters.
The parameters for the clustering are a small, interpretable set.
Appropriate values were chosen by performing scans and observing the influence of changes to the parameters on jets formed.

In section~\ref{sec:spectralmethodalgo}, 6 parameters are named:
\(\sigma_v\), \(\alpha\), \(k_\text{NN}\), \(\lambda_\text{limit}\), \(\beta\) and \stoppingdeltar{}.
While these are more parameters than in \genkt{}, for example,
we find that the parameters do not need to take precise values
to obtain good performance.
The interpretation of these parameters is as follows.
\begin{itemize}
    \item \(\sigma_v\): introduced in step~\ref{step:affinity}, this is a scale parameter in physical space.
                      The value indicates an approximate average distance for particles in the same shower,
                      or alternatively, the size of the neighbourhood of each particle.
                      It is closely tied to the stopping parameter for the \genkt{} algorithm,
                      \ktstoppingdeltar{},
                      and they both relate to the width of the jets formed.
                      It should take values of the same order of magnitude as \ktstoppingdeltar{}.
    \item  \(\alpha\): also introduced in step~\ref{step:affinity},
           this changes the shape of the distribution used to describe the neighbourhood of a particle.
           Higher values reduces the probability of joining particles outside \(\sigma_v\). In particular,  
           \(\alpha=2\) defines a Gaussian kernel.
       \item \(k_\text{NN}\): introduced in step~\ref{step:KNN}, it dictates the minimum number of non-zero affinities around each point.
           Lower values create a sparser affinity matrix, reducing noise at the potential cost of lost signal.
           Values above \(7\) are seen to have little impact.
       \item  \(\lambda_\text{limit}\): introduced in step~\ref{step:eigenvectors}, it is a means of limiting the number of eigenvectors used
           to create dimensions in the embedding space.
           Only eigenvectors corresponding to eigenvalues less than \(\lambda_\text{limit}\) are used.
           Thus, the number of dimensions in the embedding space can be increased with a larger \(\lambda_\text{limit}\). 
           However, as the eigenvalues will be influenced by the number of clear clusters available, 
           there will not be the same number of dimensions in each event.
           Values of \(0 <\lambda_\text{limit} < 1\) are sensible choices,
           see discussion in section~\ref{sec:eig_norm}.
       \item  \(\beta\): introduced in step~\ref{step:compression}, it 
          accounts for variable quality of information in the eigenvectors, as given by their eigenvalues,
        in such a way that the dimensions of the embedding spaces 
        corresponding to higher eigenvalues are compressed,
        as they contain lower quality information.
        (This is discussed in section~\ref{sec:eig_norm}.)
    \item \stoppingdeltar{}:  introduced in step~\ref{step:stoppingcondition}, it
         determines the expected spacing between jets in the embedding space.
         As the number of dimensions in the embedding space grows with increasing 
         number of clear clusters, it will not result in the same or
         similar number of clusters each time.

\end{itemize}

To investigate the behaviour of the clustering when the parameters change, scans were performed.
On a small sample of 2000 events  the clustering is performed with many different parameter choices.

With the aid of MC truth information a metric of success can be created.
For each object we wish to find (e.g., a \bthing{quark}) 
the MC truth can reveal which of the particles that are visible to the detector have
been created by that object.
In many cases, a particle seen in the detector will have been created by two objects,
such as a particle coming from a \(b\bar{b}\) pair:
in such cases both objects are considered together.
The complete set of visible particles that came from these objects could be referred to as their descendants.
The aim in jet clustering is to capture only all of the descendants in the same number of jets as there were objects that created them.
So the descendants of a \(b\bar{b}\) pair should be captured in exactly 2 jets.
The use of MC information has also been pursued in~\cite{Ju:2020tbo},
for  jets originating from a colour singlet hard particle, namely a \(W\) boson.
In addition, we will also seek to find jets emerging from systems which have a colour charge.
By allowing the descendants of groups of interacting showers to be clustered
in any configuration that results in the correct number of jets we avoid the need to associate each
descendant to one object (e.g., a~\bthing{quark}) uniquely,
which is indeed not possible when the objects in question are colour charged~\cite{Ju:2020tbo}.

    \begin{figure}[!t]
        \begin{minipage}[c]{0.6\textwidth}
            \includegraphics[width=1\textwidth]{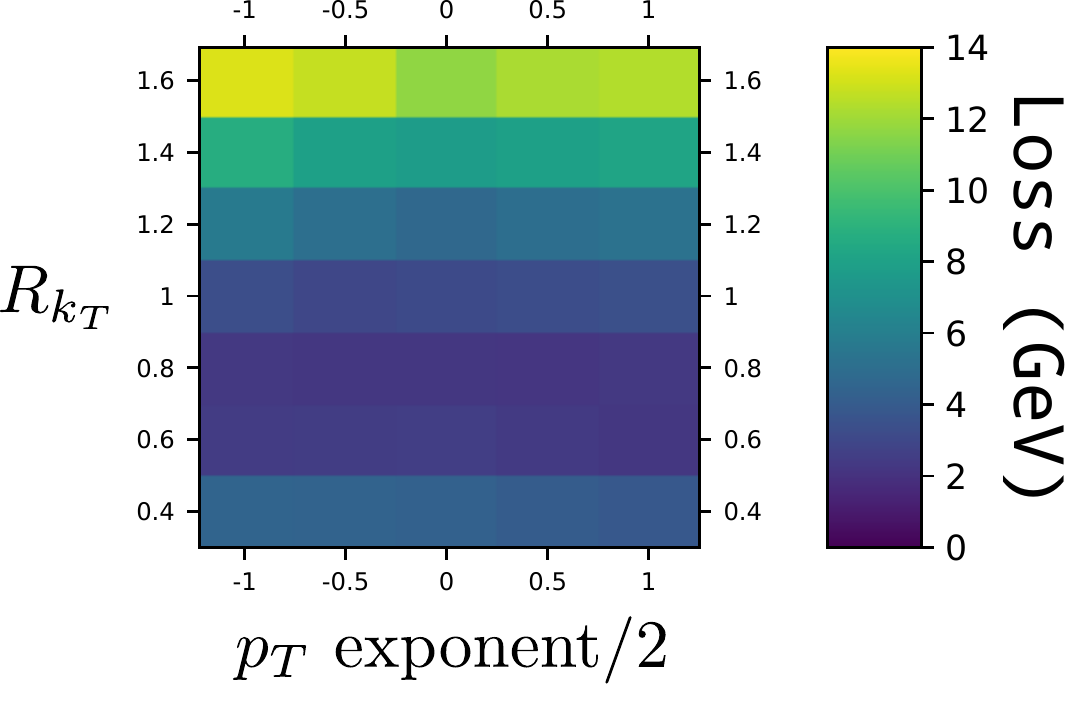}
        \end{minipage}\hfill
        \begin{minipage}[c]{0.35\textwidth}
            \caption{The \genkt{} algorithm has 2 parameters that can be varied.
                The stopping condition, \ktstoppingdeltar{}, and a multiple for the exponent of the \(p_T\) factor.
                When the exponent of the \(p_T\) factor is \(-1\) the algorithm becomes the \antikt{} algorithm.
                Here, the ``Loss'', as described in eq.~(\ref{eqn:loss}), is shown as a colour gauge for a number of parameter combinations.
             }\label{fig:scan_genkt}
        \end{minipage}
    \end{figure}    

There are two ways a jet finding algorithm can make mistakes in this task:
the first is to omit some of the descendants of the objects being reconstructed, causing the jet to have less mass than it should;
the second is to include particles that are not in the descendants of the objects being reconstructed, such as initial state radiation or particles from other objects,
causing the jet to have more mass than it should.
The effects of these mistakes might cancel in the jet mass,
but they are both still individually undesirable,
so separate metrics are made for each of them.
The first is ``Signal mass lost", the difference between the mass of the jets and the mass they would have had if all they contained were the descendants of the object being reconstructed.
The second is ``Background contamination", the difference between the mass of jets and the mass they would have if they did not contain anything but descendants of the objects being reconstructed.
A ``Loss'' function is then constructed as a weighted combination of these two,
\begin{equation}\label{eqn:loss}
\text{Loss} = \sqrt{w\,(\text{Background contamination})^2 + (\text{Signal mass lost})^2},
\end{equation}
where \(w\) is a weight used to alter the preference for suppressing ``Signal mass lost'' versus reducing ``Background contamination''.
When applying an \antikt{} algorithm, increasing \ktstoppingdeltar{} will result in lower ``Signal mass lost'', in exchange for a higher ``Background contamination''.
We have chosen to make a comparison to \(\ktstoppingdeltar{}=0.8\) as our sample dataset has well separated jets and low background.
This value of \ktstoppingdeltar{} slightly prefers suppressing ``Signal mass lost'' over ``Background contamination'',
to create the clearest mass peaks.
To make the ``Loss'' reflect this we choose \(w = 0.53\).

An example of this scan for the \genkt{} algorithm is given in figure~\ref{fig:scan_genkt}. 
It can be seen that, while good results are possible with many values of the \(p_T\) exponent,
                \ktstoppingdeltar{} must fall in a narrow range. We thus deem this choice of  stopping condition, \(\stoppingdeltar{}_{k_T}=0.8\), to be rather fine-tuned.

\begin{figure}[!t]
            \includegraphics[width=1\textwidth]{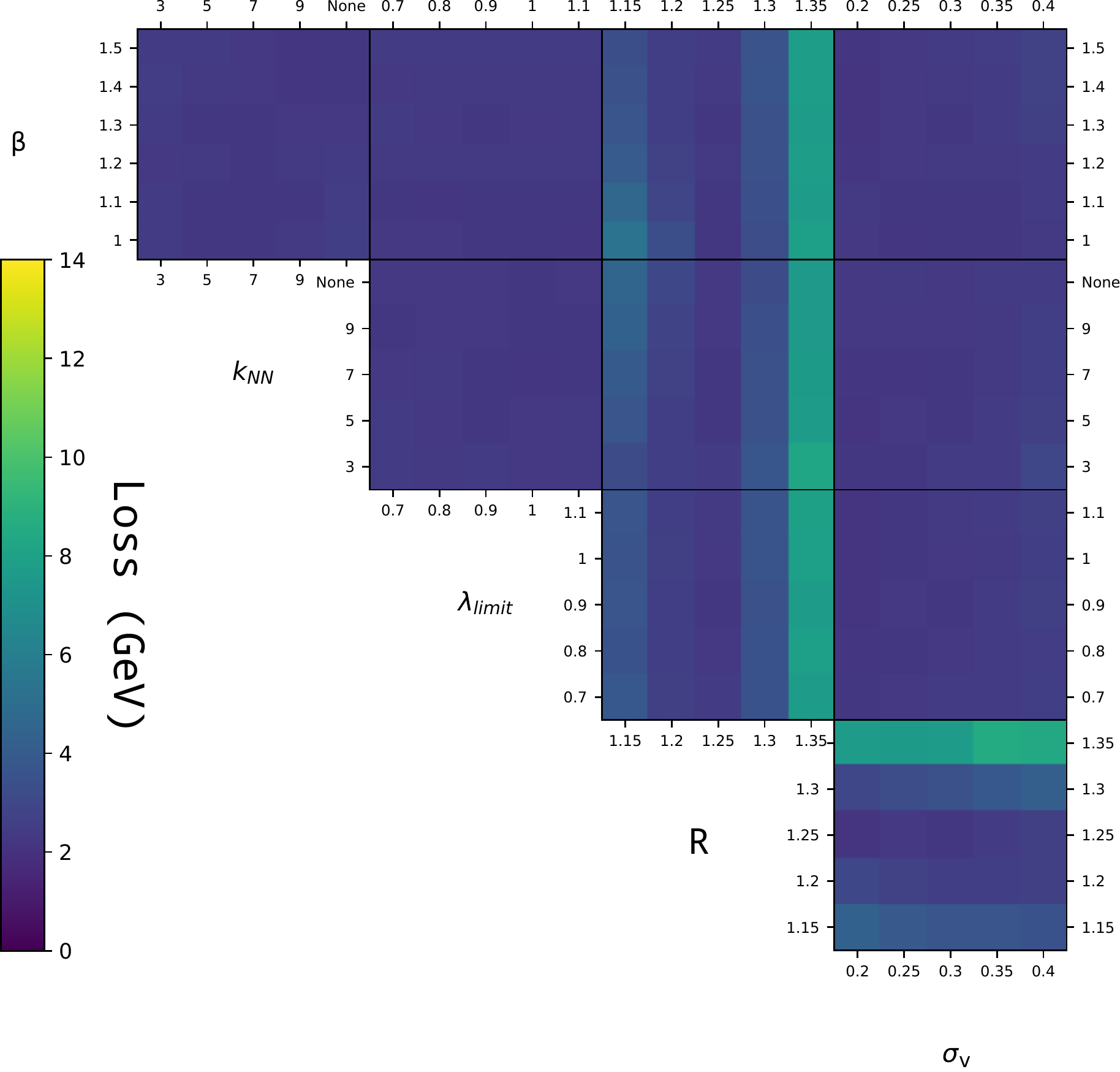}
            \caption{The spectral clustering algorithm has 6 parameters that can be varied (described in the text).
                Here, the ``Loss'', as described in eq.~(\ref{eqn:loss}), is shown as a colour gauge for reasonable  parameter ranges chosen
                either by convention (e.g., \(\alpha\) is typically \(1\) or \(2\))
                or according to physical scales (e.g., \(\sigma_v\) is of order \(0.1\)).
             }\label{fig:scan_spectral}
    \end{figure}

For \spectral{} clustering there are more than 2 variables to deal with, 
so a set of two dimensional slices are extracted. 
These slices have been chosen to include the best performing combination.
They are plotted in figure~\ref{fig:scan_spectral} with the same colour scale as figure~\ref{fig:scan_genkt},
to allow for direct comparison.
    As can be seen in figure~\ref{fig:scan_spectral}, the parameters choices are not fine-tuned,
    as many values can be chosen to achieve good results.
    For example, it can be seen that some parameters, such as \(\alpha\), \(k_\text{NN}\), \(\beta\)
    and \(\lambda_\text{limit}\), are relatively unconstrained,
    yielding  good results for a wide range of numerical choices.
    Even when \stoppingdeltar{} and, especially, \(\sigma_v\) yield some large signal ``Loss'',
    say, for \(R=1.35\) and \(\sigma_v=0.4\), this happens in very narrow ranges. 
    For definiteness, the
    parameters used in the remainder of this work are \(\alpha=2\), \(k_\text{NN}=5\), \(\stoppingdeltar{} = 1.26\), \(\beta = 1.4\), \(\sigma_v = 0.15\) and \(\lambda_\text{limit} = 0.4\).

    \subsection{Particle data}\label{sec:particle_data}

To evaluate the behaviour of the spectral clustering method four datasets are used,\footnote{The first two uses a 2-Higgs Doublet Model (2HDM) setup as described in ref.~\cite{Chakraborty:2020vwj} while the last two are purely Standard Model (SM) processes. Notice that all unstable objects are rather narrow, including the Beyond the SM (BSM) Higgs states \cite{Moretti:1994ds,Djouadi:1995gv}, so that we have neglected interference effects with their irreducible backgrounds.} all produced for the LHC.

    \begin{enumerate}
        \item 
    \underbar{Light Higgs}: A SM-like Higgs boson with a mass \(125\) GeV decays into two light Higgs states with mass  \(40\) GeV,
    which in turn decay into \beau{}\bbar{} quark pairs.
    That is, the process is \(gg,q\bar q \rightarrow H_{125\,\text{GeV}} \rightarrow h_{40\,\text{GeV}} h_{40\,\text{GeV}} \rightarrow \beau \bbar \beau \bbar\), simulated at Leading Order (LO). (Here, $m_b=4.75$ GeV for the $b$-quark pole mass.) 

\item \underbar{Heavy Higgs}:  A heavy Higgs boson with a mass \(500\) GeV decays into two SM-like Higgs states with mass  \(125\) GeV,
    which in turn decay into \beau{}\bbar{} quark pairs.
    That is, the process is \(gg,q\bar q \rightarrow H_{500\,\text{GeV}} \rightarrow h_{125\,\text{GeV}} h_{125\,\text{GeV}} \rightarrow \beau \bbar \beau \bbar\), simulated at LO.

\item \underbar{Top}:  A $t\bar t$ pair decays semileptonically, i.e.,  one \(W^\pm\) decays into a pair of quark jets $jj$ and the other into a lepton-neutrino pair $\ell\nu_\ell$ ($\ell=e,\mu$).
        That is, the process is \( gg,q\bar q \rightarrow t \bar{t} \rightarrow b\bar b W^+  W^-\to b\bar b jj \ell\nu_\ell\), simulated at LO. (Note that, here, $m_t=172.6$ GeV and $m_{W}=80.4$ GeV.)
        
    \item \underbar{3-jets}:  For the purpose of checking {IR safety} we have used 3-jet events,  this being a rather simple configuration where IR singularities could be observed. 
        That is, the process is $pp\to jjj$, simulated at both LO and Next-to-LO (NLO).

    \end{enumerate}

    Using MadGraph~\cite{Alwall:2011uj} to generate the partonic process and Pythia~\cite{Sjostrand:2014zea} to shower, ${\cal O}(10^5)$ events for each of these processes are generated.
    A full detector simulation is not used, instead, cuts on the particles are imposed to approximate detector resolution, as detailed below. 
    
    The Center-of-Mass (CM) energy used is \(\sqrt{s}=13 \) TeV.
    Each event also contains (hard) Initial State Radiation (ISR) and soft QCD dynamics from beam remnants, i.e., the Soft Underlying Event (SUE).
    Two versions of each dataset are produced, one with Multi-Parton Interactions (MPIs) but not pileup,
    and the second with pileup and MPI.
    The simplistic case of clustering on signal without pileup is explored to start with.
    Pileup creates substantial additional noise that greatly complicates clustering jets.  While various pileup mitigation and jet grooming techniques might be applied to the data, it is beyond the scope of this work.
    However, it is still interesting to see the unadulterated behaviour of a clustering algorithm in the presence of pileup: we explore this in a later section.
    
    To simulate  pileup, \(10^5\) minimum bias events are generated in Pythia.
    For each signal event, a number of pileup events are merged into the event:  this number is drawn from a Poisson distribution with mean $\lambda=50$.  These pileup events are introduced with a vertex displacement of  \(\lesssim \pm0.1\mathrm{mm}\), as described in~\cite{pileup_mitigation2019}.

    Each of these datasets requires different cuts, both at the particle level, to simulate detector coverage, and at the jet level, to select the best reconstructed events.
    The cuts on each dataset are as follows.
    \begin{enumerate}
        \item The reconstructed particles are required to have
            pseudorapidity \(|\eta|< 2.5\) and transverse momentum \(p_T > 0.5\) GeV.
            These cuts are likely to remove the majority of the radiation from beam remnants
            and reduce ISR.
            The \bthing{jets} are required to have \(p_T > 15\) GeV, which is possibly lower than is realistic \cite{Chakraborty:2020vwj},
            but it leaves a larger number of events to compare the behaviour of jet clustering algorithms.

        \item  The reconstructed particles are required to have
             \(|\eta|< 2.5\) and \(p_T > 0.5\) GeV.
            The $b$-jets are required to have \(p_T > 30\) GeV, which is realistic for efficient $b$-tagging performance and further reduces ISR and the SUE.
            As the average jet \(p_T\) is higher we can afford this higher \(p_T\) cut.
            
        \item The reconstructed particles are required to have
             \(|\eta|< 2.5\) and \(p_T > 0.5\) GeV.
            The event is required to have  \(p_{T}^{\text{miss}} > 50\) GeV,
            where \(p_{T}^{\text{miss}}\) is the missing transverse momentum due to 
            the neutrino.
            The lepton in the event must have  \(|\eta|< 2.4\).
            If the lepton  is a muon then its \(p_T\) must be \(>  55\) GeV.
            If the lepton  is an electron and it is isolated (as defined in~\cite{Sirunyan:2018fpa}) then its \(p_T\) must be \(> 55\) GeV, if it is not isolated then \(p_T > 120\) GeV.
            The reconstructed jets must have \(p_T > 30\) and \(|\eta|< 2.4\).
            Finally, the lepton must be separated from the closest jet by at least
            \(\sqrt{\Delta\eta^2 + \Delta \phi^2} > 0.4\) or
            \(p_{T}^{\text{relative}} > 40\) GeV.
            These cuts are copied from~\cite{Sirunyan:2019rfa}.
        \item The only restriction on the particles is through the pseudorapidity,  \(|\eta|< 2.5\).
            There are no cuts on the jets. We adopt this unrealistic condition in order to explore issues of IR safety, since these are emphasised at low \(p_T\).

    \end{enumerate}

    For all datasets with pileup, any charged tracks that originate from a vertex that
    are displaced by at least \(75\) $\mu$m from the primary vertex are removed. 
	This removes the majority of charged pileup tracks in the dataset, leaving mostly neutral tracks.

    The Higgs boson cascade datasets have the desirable property of creating \bthing{jets} with different kinematics: while in case 1 we may expect some slim  jets (as on average they are rather stationary, because of the small mass difference between $H_{125\,{\rm GeV}}$ and $h_{40\,{\rm GeV}}$)
in case 2 we may see mainly fat jets (owing to the boost provided by the large mass difference between $H_{500\,{\rm GeV}}$ and $h_{125\,{\rm GeV}}$).
Mass reconstruction requirements for the \underbar{Light Higgs} and \underbar{Heavy Higgs} follow the same logic.
In order to reconstruct a Higgs boson decaying directly to a pair of \bthing{quarks}, we require a separate jet tagged by each \bthing{quark}, that is, two jets are required, each tagged by a \bthing{quark} from that Higgs state.
To reconstruct a Higgs boson that decays into a pair of (child) Higgs particles,
we require both child Higgs bosons to have been reconstructed,
that is, all four \bthing{jets} are found.

In the case of the \underbar{Top} events
three masses can be reconstructed from jets, the hadronic \(W\),
the hadronic top and the leptonic top.
The hadronic \(W\) is reconstructed if both of the quarks it decayed into have tagged jets: 
they are permitted to tag the same jet, so the hadronic \(W\) can be reconstructed from one or two jets.
The hadronic top is reconstructed if the \bthing{quark} from it has tagged a jet, so the correct \bthing{jet} is required in addition to the requirements on the \(W\).
The leptonic top is reconstructed if the \bthing{quark} from the top decay tags a jet and the missing momentum calculation which reconstructs the leptonic \(W\) yields a real mass.
If the mass calculation for  the leptonic \(W\) yields two real masses, the one closest to the true \(W\) mass is selected.

We now proceed to compare spectral to \antikt{} and CA clustering. We start from testing IR safety of the former, while this is a well-known feature of the latter two.
We will then move on to study Higgs boson and top quark events.

\subsection{Checking sensitivity to IR behaviour}\label{sec:IRmethod}
The algorithm itself will be IR safe due to the \(s(t)_{i,j}\) factors.
These factors force any soft or collinear particles to merge first
and ensure that they do not alter the size, \(z(t)_j\), of the pseudojets.

This has been verified by taking toy datasets, which are varied by adding soft particles
or splitting particles in a collinear fashion.
The clustering on these datasets never alters under these variations, 
until the limits of computational precision are reached.

    As the environment required for clustering on MC data is already set up,
    it is rather efficient for this study to offer proof
    that in practice the algorithm is not sensitive to
    IR considerations in simulated data.
    This can be done by showing that an IR sensitive variable, for example,
    the jet thrust spectrum,
    is stable between a LO dataset with no IR singularities and a NLO
    one which will instead contain IR singularities.
    This is a very important property, as the algorithm must not
    be modified by any approximation used for the IR limit in MC simulation.

    Showing the jet thrust at LO and NLO for a particular configuration,
    that is, a particular selection of clustering parameters,
    would allow a comparison that would highlight any differences caused by IR sensitivity.
    This will be done for illustrative purposes,
    however, since even an IR unsafe algorithm, such as the iterative cone one~\cite{Cacciari:2008gp},
     has some configurations for which these singularities are avoided.
    To provide a more global view, a scan of parameter configurations must be compared.
    Thus, for an unsafe algorithm (such as the iterative cone) the unsafe configuration
    will be found.
    It would be cumbersome to compare all these thrust distributions by eye, however.
    Instead, we introduce a summary statistic representing the divergence between two distributions,
    the Jensen-Shannon score~\cite{Lin:1991zzm}.

    The Jensen-Shannon score is a value computed between two distributions that increases in magnitude the more these distributions differ.
    It is a symmetrised variant of the Kullback-Leibler divergence~\cite{Lin:1991zzm}.
    The Kullback-Leibler divergence between probability densities \(p\) and \(q\) can be written as
    \begin{equation}
    D_\text{KL} (p | q) = \int^{\infty}_{-\infty} p(x) \log\left(\frac{p(x)}{q(x)}\right) dx,
\end{equation}
    from which the Jensen-Shannon divergence can be written as
    \begin{equation}
    D_\text{JS}(p, q) = \frac{1}{2}D\left(p | \frac{1}{2}(p + q)\right) + \frac{1}{2}D\left(q | \frac{1}{2}(p + q)\right).
\end{equation}
    Here, \(D_\text{JS}\) treats \(p\) and \(q\) symmetrically and will grow as they become more different.
    The spectrum of Jensen-Shannon scores will be plotted for a known IR safe clustering algorithm, \genkt{},
    a known unsafe clustering algorithm, iterative cone, and the \spectral{} algorithm.
    If the Jensen-Shannon scores for \spectral{} are consistently small,
    then it is IR safe.

    \FloatBarrier
    \section{Results}

\begin{figure}[htp]
    \begin{center}
    \includegraphics[width=\textwidth]{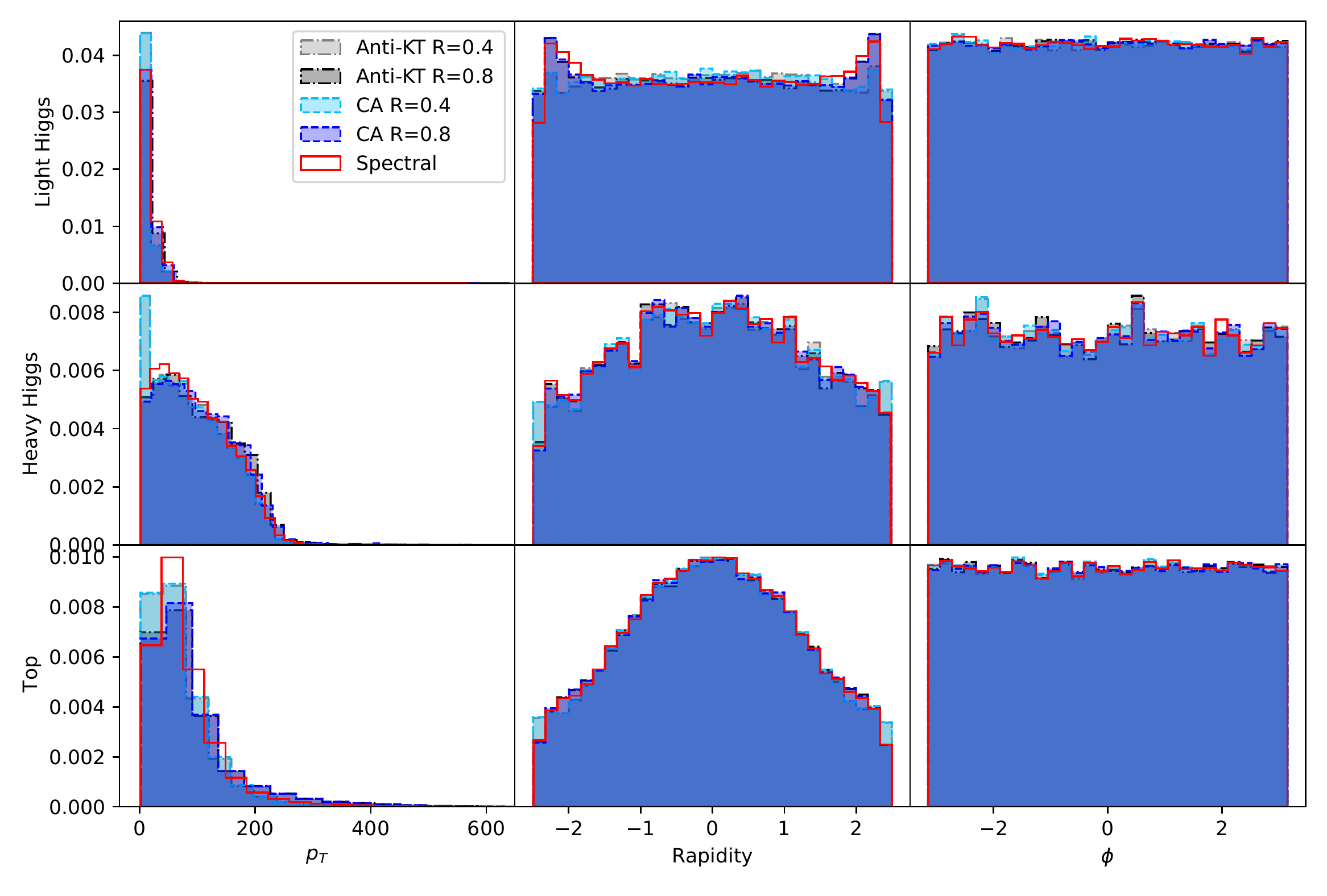}
        \caption{Basic jet variables for each of the analysis datasets and three clustering algorithms.
            In the first column there are some noticeable differences in the transverse momentum.
            In the second column the rapidity shows that
            the algorithms cluster jets at the edge of the barrel slightly differently.
            In the third column the barrel angle shows no noticeable changes.
        }\label{fig:kinematics}
\end{center}
\end{figure}

Before the behaviour of the algorithms is analysed, some plots of kinematic variables are shown
in figure~\ref{fig:kinematics}.
It can be seen that the algorithms do not greatly differ on the kinematics of the events.
Spectral creates jets with similar kinematic features to the CA and \antikt{} algorithms.
In particular, \spectral{} clustering does not appear to sculpt any distributions in any of the datasets involving Higgs bosons and top (anti)quarks.

\subsection{IR safety}
\begin{figure}[htp]
\includegraphics[width=\textwidth]{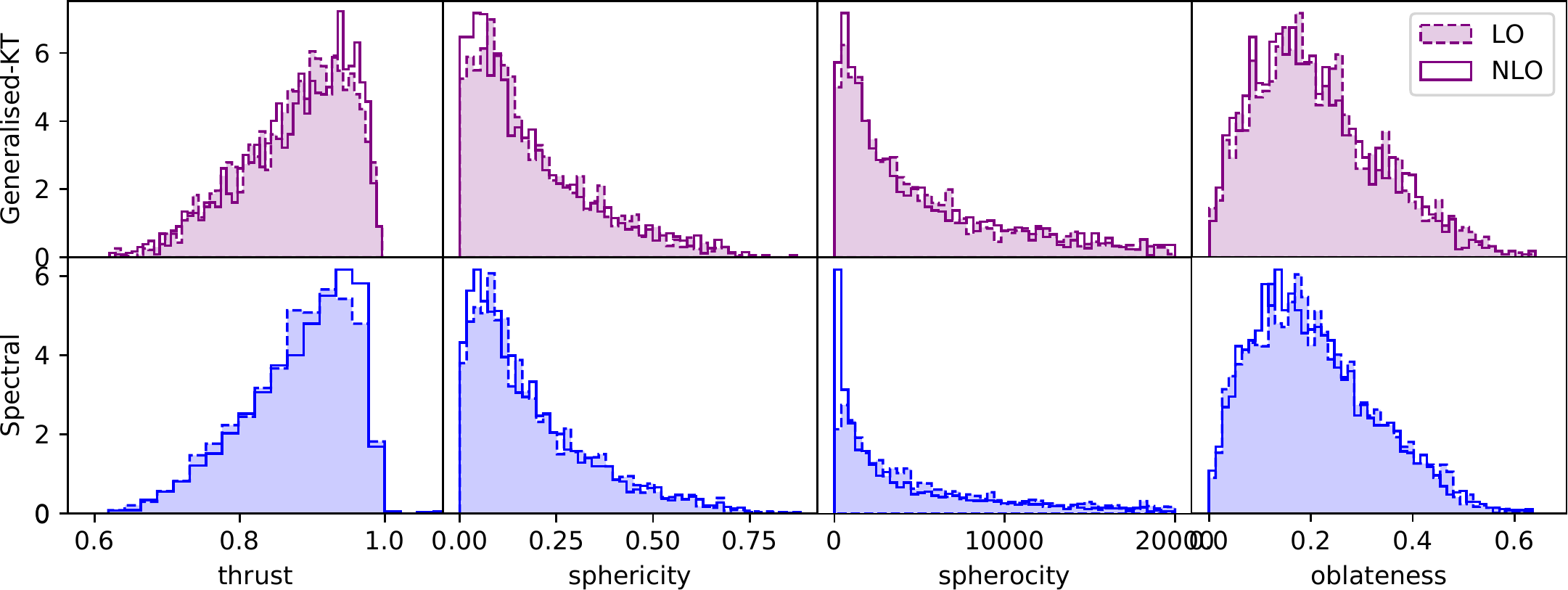}
    \caption{Spectra for jet properties created with LO and NLO datasets.
             The \(4\) jets with highest \(p_T\) from each event are used in aggregate as an average to 
             form these plots.
             The columns from left to right are: the jet
             thrust, sphericity, spherocity and oblateness.
             Algorithms where configured (i.e., the settings of \stoppingdeltar{} chosen)
             to give sensible results on
             this dataset, therefore distributions may not represent worst case scenarios.
         }\label{fig:IRC_singles2}
\end{figure}

Shape variables (see the QCD section of ref.~\cite{Altarelli:1989hv} for a useful review), such as jet thrust, sphericity, spherocity and oblateness,  are sensitive to IR divergences.
For each configuration of the clustering algorithm we expect an IR safe algorithm to present a stable transition
in a shape variable from the LO to NLO datasets, as significant
changes in the spectra would indicate sensitivity to soft and collinear radiation.
The clustering and evaluation here is done using the \underline{3-jets} dataset, as described in section~\ref{sec:particle_data}.
Shape variables are calculated from the total momentum of the 4 jets with highest \(p_T\) in each event.
This comparison is made in figure~\ref{fig:IRC_singles2}.
It can be seen in this figure that little difference exists between \genkt{} and \spectral{} clustering, so as to reinforce that they are both IR safe.

\begin{figure}[htp]
\includegraphics[width=\textwidth]{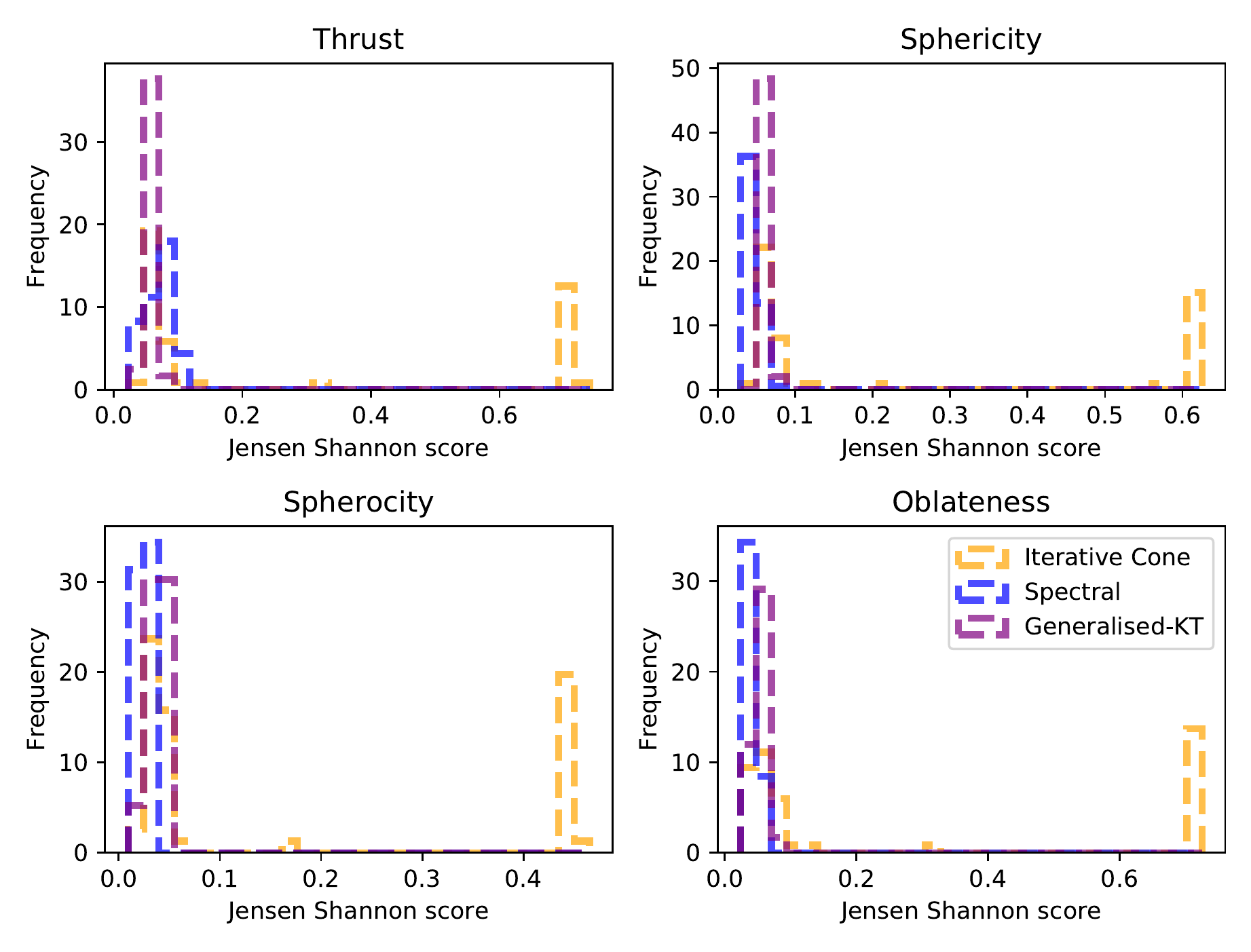}
    \caption{
        Histograms evaluating IR safety from each jet shape variable.
        Each count is a  Jensen-Shannon score between a probability density of the
        jet shape variable from LO and NLO data.
        Counts at low values indicate insensitivity to IR differences between the LO and NLO data,
        thus IR safety.
     }\label{fig:unnormedJS}
\end{figure}

However, this method of establishing IR safety only looks at one parameter configuration and could be accused of cherry-picking.
As described in section~\ref{sec:IRmethod}, this can be systematically compared for many parameter configurations by calculating a Jensen-Shannon
score for each LO and NLO pair of jet shape spectra.
If the Jensen-Shannon metric is low, then the two distributions are similar and appear IR safe.
To further clarify the result we include an algorithm known to be IR unsafe, the \itercone{} algorithm, as intimated.
The spectral method produces Jensen-Shannon scores very similar to \genkt{} methods.
Only the iterative cone algorithm produces high Jensen-Shannon scores thus indicating significant changes between the LO and NLO spectra.
This can be seen in figure~\ref{fig:unnormedJS}.

From these two figures it is clear that \spectral{} clustering
is not sensitive to IR effects between LO and NLO data.
It behaves at least as well as \genkt{} methods.
This contrasts with the \itercone{} algorithm, for which the jet shape spectra at LO and NLO 
differ significantly for many configurations.
This is not unexpected as  the \spectral{} clustering algorithm 
has been specifically tailored for IR safety. We have thus evidenced this safety by simulation.

\subsection{Mass peak reconstruction without pileup}\label{sec:without}
In this section, the \antikt{} and CA algorithms
with jet radius \(\ktstoppingdeltar{} = 0.4\) and \(\ktstoppingdeltar{} = 0.8\)
are compared to the \spectral{} algorithm specified in section~\ref{sec:spectralmethodparam}.
The jets are tagged using MC truth.
To start with, all datasets are considered without pileup.  We introduce 
 a tagging distance metric 
\[d_{\mathrm{tag}}:=\sqrt{(y_\text{quark tag} - y_\text{jet})^2 + \delta(\phi_\text{quark tag}, \phi_\text{jet})^2},\]
where \(\delta(\phi_\text{quark tag}, \phi_\text{jet})\) is as defined in eq.~(\ref{eqn:cyclic_phi}).
The identity of the \bthing{quarks} created by a signal particle (either a Higgs boson or a top (anti)quark) is used to tag the closest jet within $d_{\mathrm{tag}}\leq 0.8$.
In the case of a \(W\) decay, the procedure is the same applied to light quark states.
From this point on, only jets tagged this way are considered.

\begin{figure}[htp]
    \begin{center}
        \begin{minipage}[c]{0.46\textwidth}
        \includegraphics[width=1.\textwidth]{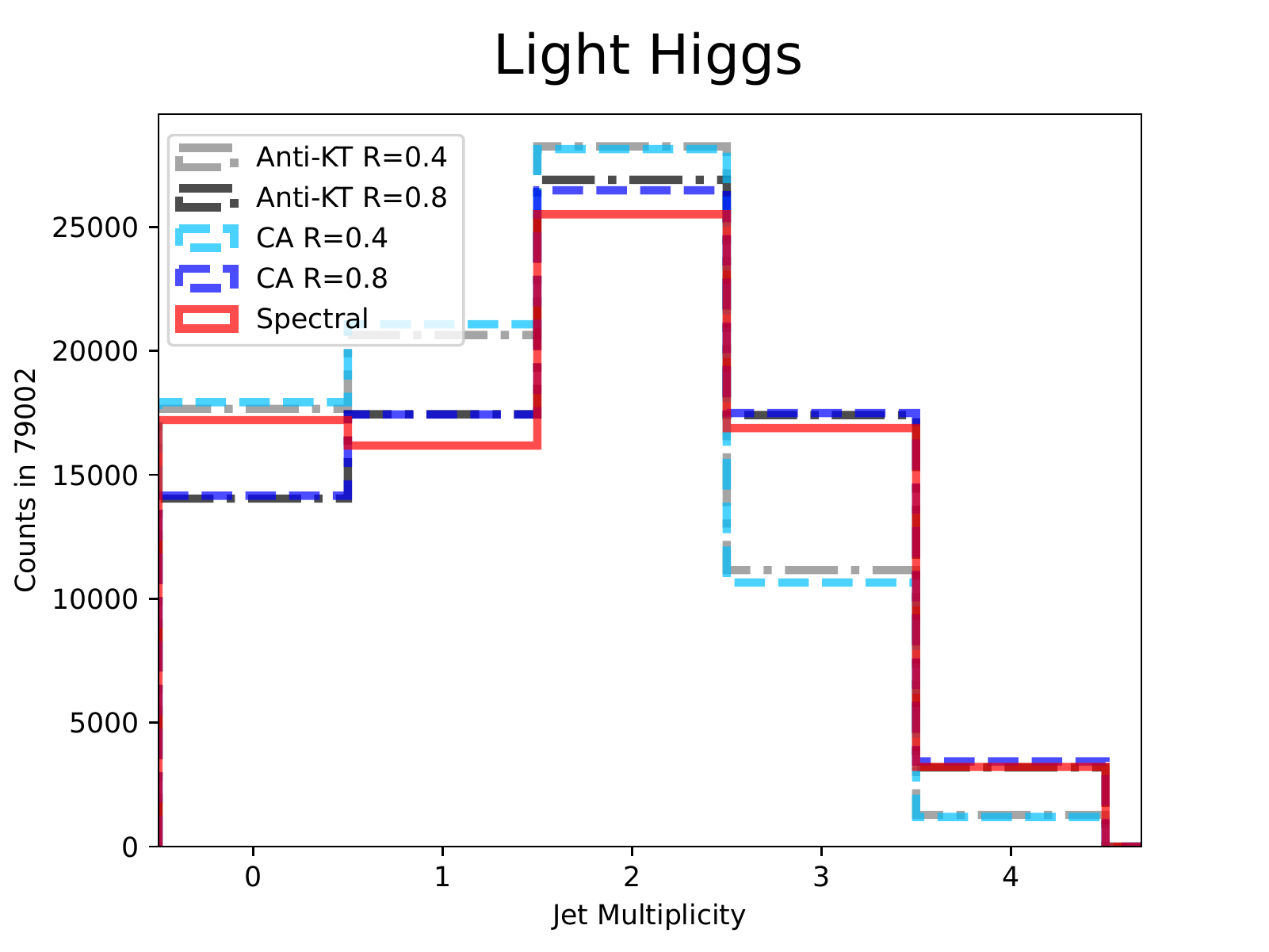}\\
        \includegraphics[width=1.\textwidth]{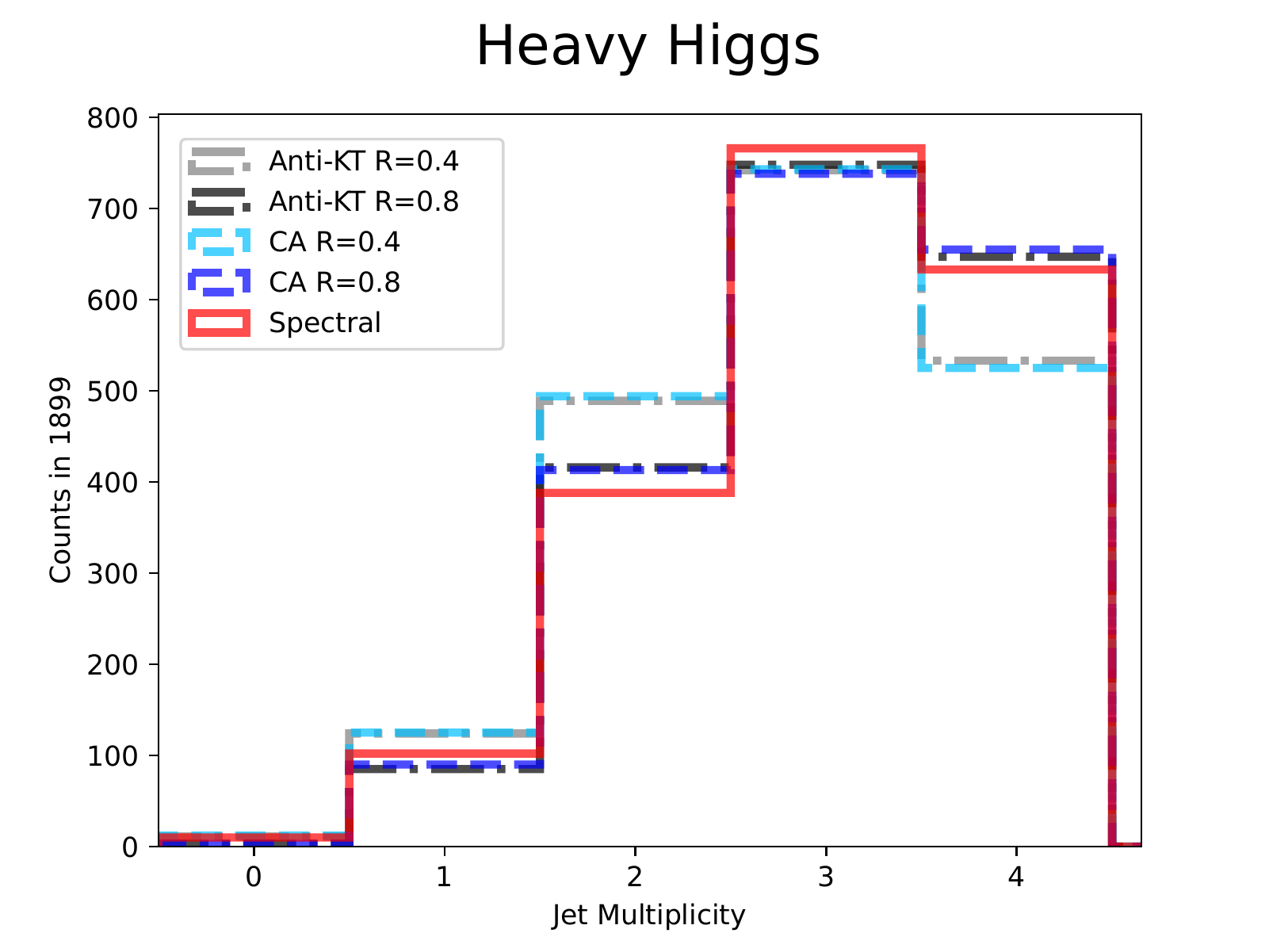}
        \end{minipage}\hfill
        \begin{minipage}[c]{0.46\textwidth}
        \includegraphics[width=1.\textwidth]{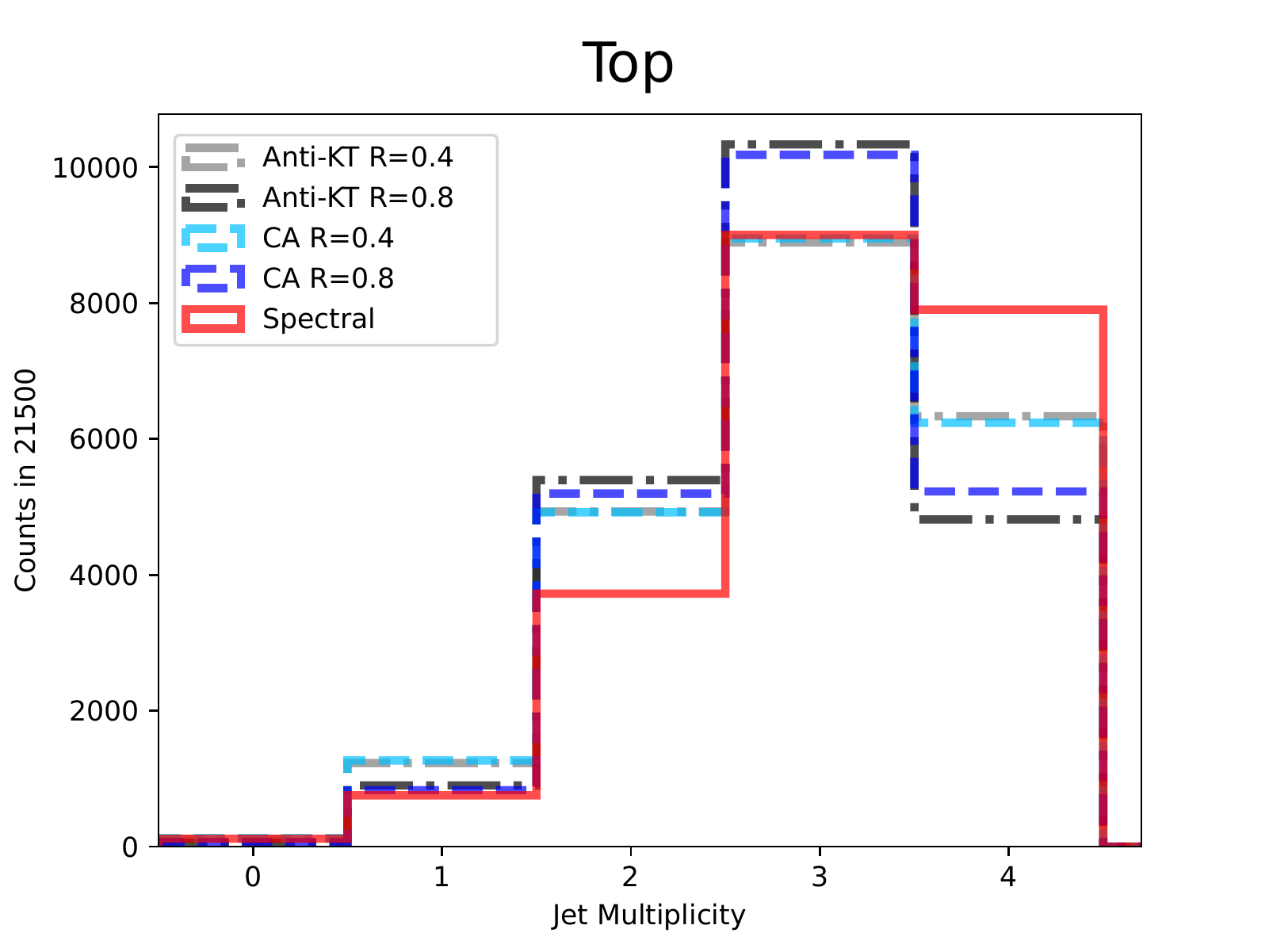}
    \caption{Jet multiplicities for the anti-$k_T$ and CA (for two \(R_{k_T}\) choices) and \spectral{} clustering algorithms on the \underline{Light Higgs}, \underline{Heavy Higgs} and \underline{Top} MC 
 samples. For all such datasets, the hard scattering produces  4 partons in the final  state, so maximising a multiplicity of 4 jets indicates good performance.   
    }\label{fig:multiplicity}
        \end{minipage}
    \end{center}
\end{figure}

Jet multiplicities, that is, the number of reconstructed jets found per event, are given for the anti-$k_T$, CA and spectral clustering algorithms.
These can be seen for the first three datasets described in section~\ref{sec:particle_data} in figure~\ref{fig:multiplicity}. Herein, it
 is seen that \spectral{} clustering produces the best multiplicity (i.e., most events where 4 jets are found) for \underline{Top} events while for 
         the \underline{Light Higgs} and \underline{Heavy Higgs} MC samples  
        it creates a multiplicity closer to that of \antikt{}/CA\footnote{
        CA and \antikt{} being very similar in behaviour.} 
        with \(\ktstoppingdeltar{} = 0.4\) 
        than \(\ktstoppingdeltar{} = 0.8\), the first of these being the best performer of the two.
        This study provides evidence that spectral clustering, unlike anti-$k_T$, adapts to the different final states without having to adjust its parameters. The anti-$k_T$ algorithm suggests 0.4 to be the best choice for all datasets, but this is in tension with the fact that different masses from different datasets do require the anti-$k_T$ parameters to be adjusted, as we shall now see. 
Mass peaks are constructed from the reconstructed jets as well as, for the top sample only, from the lepton and neutrino.
Again, the \antikt{} and CA results with \(\stoppingdeltar{}_{k_T} = 0.4\) and \(0.8\) are given for comparison.

\begin{figure}[htp]
    \includegraphics[width=1.\textwidth]{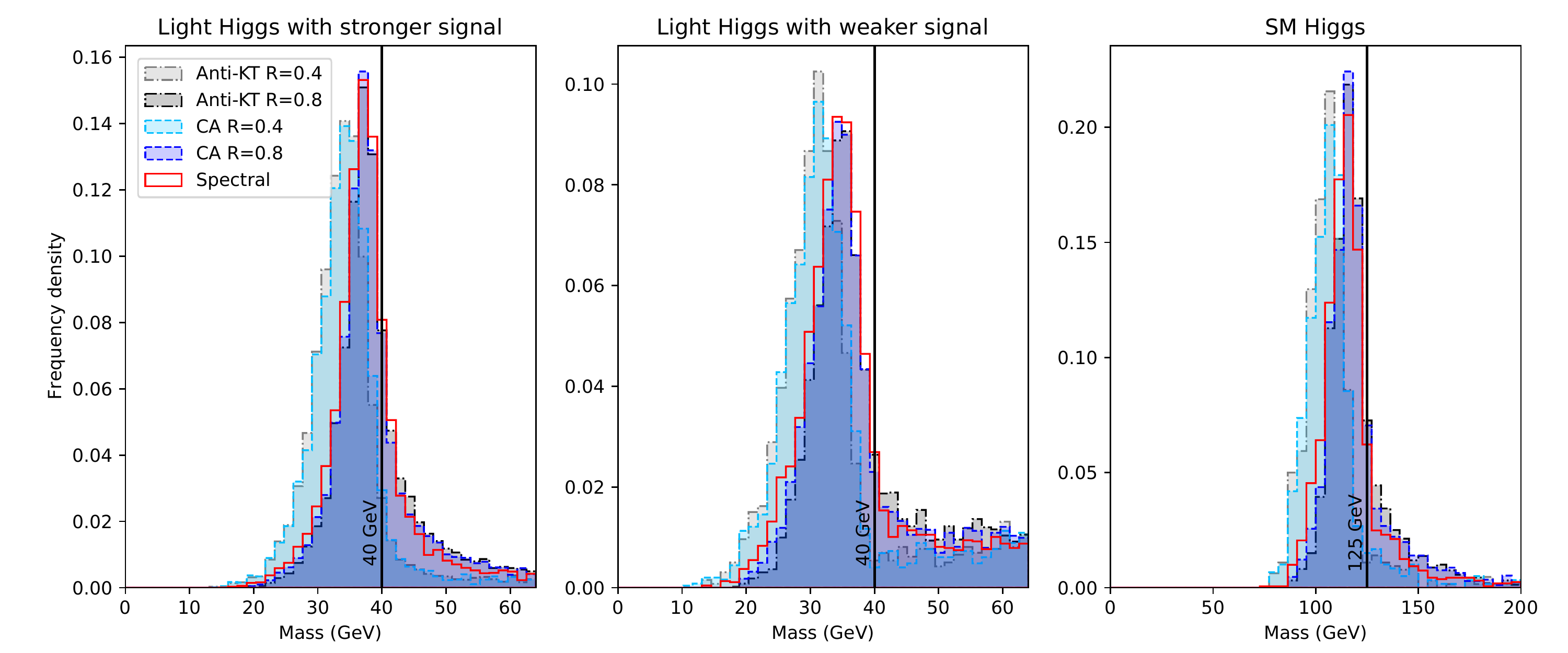}
    \caption{Three mass selections are plotted for the \underline{Light Higgs} dataset. From left to right we show: the invariant mass of the $4b$-jet system, of the $2b$-jet system with heaviest invariant mass and of the $2b$-jet system with lightest invariant mass (as defined in the text).   Three jet clustering combinations are plotted as detailed in the legend.
        The spectral clustering algorithm is consistently the best performer in terms of the narrowest peaks being reconstructed and comparable to \antikt{}/CA with \(\ktstoppingdeltar{} = 0.8\) in terms of their shift from the true Higgs mass values, with \antikt{}/CA with \(\ktstoppingdeltar{} = 0.4\)  being the outlier. 
    }\label{fig:best_correct_h_allocation}
\end{figure}

In figure~\ref{fig:best_correct_h_allocation} three selections are plotted for the \underline{Light Higgs} MC sample. We show events where all four \bthing{jets}
are combined into the total invariant mass of the event, thus reconstructing the mass of the SM Higgs boson.
Each event also contains two light Higgs states, though. These are differentiated by the mass of the particles (generated by them) that pass the particle cuts,
as follows. The light Higgs boson reconstructed from the $2b$-jet system with more mass visible to the detector is called the ``Light Higgs with stronger signal''
while the one reconstructed  with less mass visible in the detector is called the ``Light Higgs with weaker signal''.
The correct jets for each Higgs mass reconstruction are identified using MC truth,
so the correct pairings are always made. (If two such dijet systems are not found the event is not included in the plots).
Altogether, it can be seen that spectral clustering forms the best peaks,
narrow and close to the correct mass. In fact, its performance
is comparable to that of anti-$k_T$ with \(R_{k_T}=0.8\) and is clearly better than the 0.4 option. 
When the parameters for spectral clustering were chosen,
they were designed to minimise a loss
that was based on the behaviour of CA with \(R_{k_T}=0.8\) on this \underline{Light Higgs} dataset.\footnote{See section~\ref{sec:spectralmethodparam}.}
Given this, the similarity of the mass peaks is not surprising.
It will be more interesting to see how the algorithm treats a different dataset.

\begin{figure}[htp]
    \includegraphics[width=1.\textwidth]{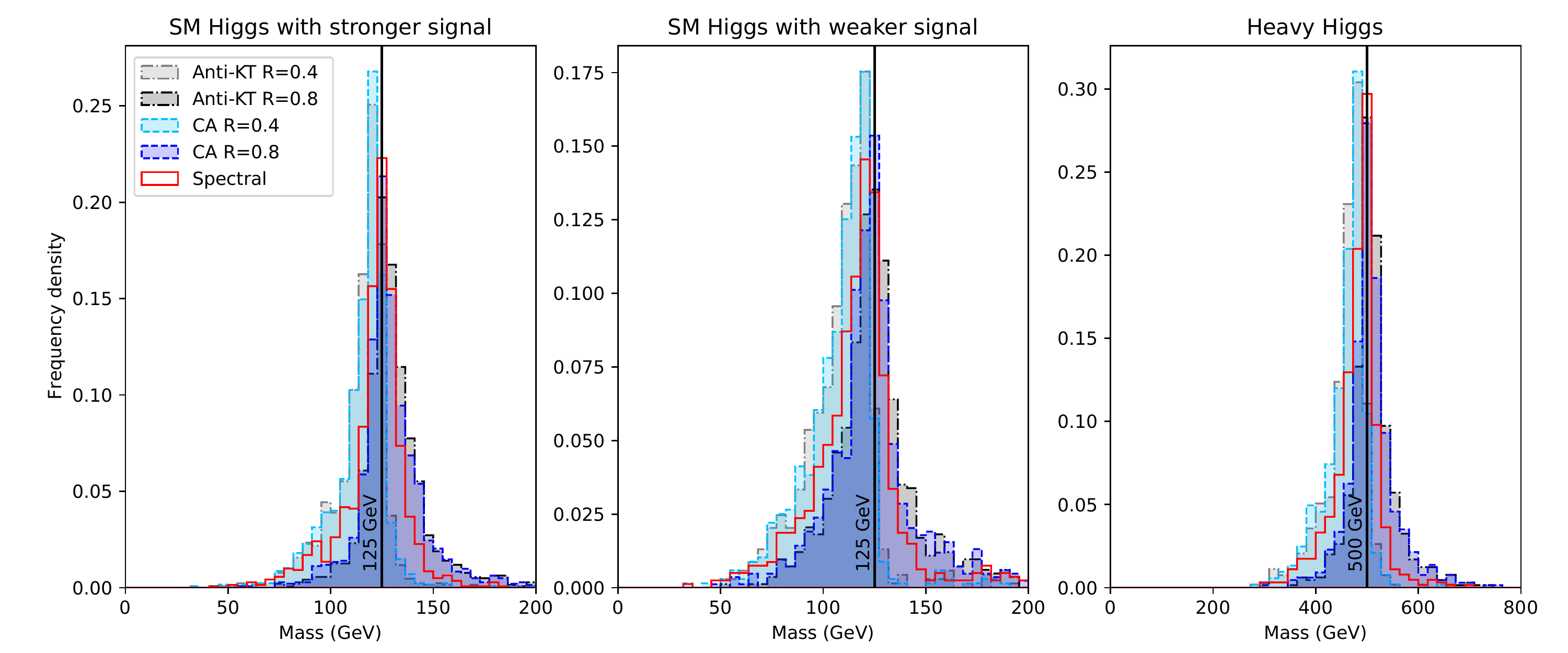}
    \caption{Same as figure~\ref{fig:best_correct_h_allocation} for the \underline{Heavy Higgs} dataset. Here, the performance of the spectral clustering and \antikt{}
        (with both 0.4 and 0.8 as jet radii) clustering algorithms  is much closer to each other. 
}\label{fig:heavy_correct_mass_peaks}
\end{figure}

In 
figure~\ref{fig:heavy_correct_mass_peaks} the exercise is repeated for the \underline{Heavy Higgs} MC dataset.
All the parameters of \spectral{} clustering are the same as in the \underline{Light Higgs} MC sample yet we note that 
its performance is still excellent, with very sharp peaks at the correct masses, although the three clustering algorithms are overall much closer in performance.
Recall that, in figure~\ref{fig:multiplicity},
it was seen that spectral clustering achieved better multiplicity than \antikt{} or CA with \(\ktstoppingdeltar{} = 0.8\) on this dataset. Furthermore, 
while the multiplicity of \antikt{} and CA with \(\ktstoppingdeltar{} = 0.4\) is about equivalent,
the location of all Higgs mass peaks for \antikt{} with 
\(\ktstoppingdeltar = 0.4\) is slightly worse.
So, we are again driven to conclude that spectral clustering is probably the best performer overall with the added benefit of not requiring any adjustment of its parameters to achieve this. 

\begin{figure}[htp]
    \includegraphics[width=1.\textwidth]{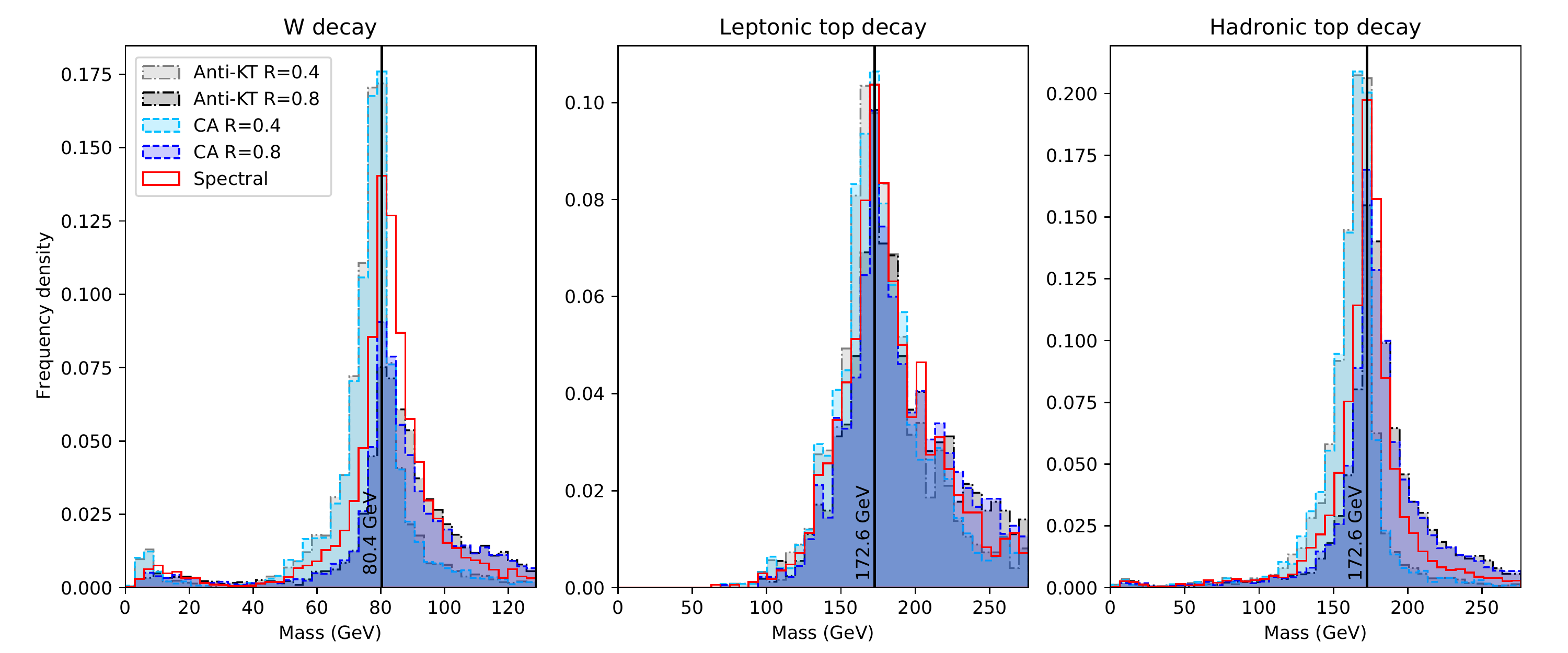}
    \caption{Three mass selections are plotted for the \underline{Top} dataset. From left to right we show: the invariant mass of the light jet system, of the reconstructed leptonic $W$ (as  described in the text) combined with a $b$-jet and of the hadronic $W$ combined with the other $b$-jet. Three jet clustering combinations are plotted as detailed in the legend.
The spectral clustering algorithm  consistently outperforms  anti-$k_T$ with jet radius 0.8 and is slightly worse than the anti-$k_T$/CA one with \(R_{k_T}=0.4\), but only in terms of sharpness, not of location of the mass peak.
    }\label{fig:top_correct_mass_peaks}
\end{figure}    

Finally, in figure~\ref{fig:top_correct_mass_peaks}, the $W$ and $t$ mass peaks for semi-leptonic $t\bar t$ decays are shown.
Three mass reconstructions are given. The hadronic \(W\) is reconstructed from the jets that come from the quarks it decayed to.
Correct decisions about which quarks correspond to which particle in the hard process are made by using information in the MC:
this is to prevent performance evaluation of clustering to be confounded by mismatching.
To tag a jet with a quark we use the tagging distance measure $d_{\mathrm{tag}}$.
The \(W\) will always decay to a pair of quarks, which may be captured in one jet or separate jets.
If either of the these quarks are too far away from the closest jet to tag it,
that is \(d_{\mathrm{tag}} > 0.8\),
then it is not associated with any jet and the hadronic \(W\) is not reconstructed.
The mass of the hadronic top is then reconstructed in events where the hadronic \(W\) could be reconstructed and the \bthing{jet}
from the hadronic top is also found.
The leptonic top is then reconstructed in events where a \bthing{jet} from the top is combined with the reconstructed $W$ which decays leptonically.
The leptonic reconstruction of the $W$ uses the momentum of the electron $p_\ell$, the missing transverse momentum $p_T^{\rm miss}$ (identified with that of the neutrino)
and the longitudinal neutrino momentum ($p_L^\nu$, which is unknown) in a quadratic equation, $(p_\ell+p_T^{\rm miss}+p_L^\nu)^2=m_W^2$, of which only the real solutions are plotted.  In this case, it can be seen that \spectral{} clustering is adapting to jets of a different radius. In fact, 
while before its behaviour had mostly resembled anti-$k_T$ with \(\ktstoppingdeltar{} = 0.8\), 
it has now moved closer to the case with \(\ktstoppingdeltar{} = 0.4\).
(Semi-leptonic top events would typically be processed using anti-$k_T$ with \(\ktstoppingdeltar{} = 0.4\).)
The peaks of \spectral{} clustering are not quite as narrow as those from anti-$k_T$ with \(\ktstoppingdeltar{} = 0.4\),
but they improve on \(\ktstoppingdeltar{} = 0.8\) and their  location is substantially correct.
Furthermore, the multiplicity obtained by \spectral{} clustering on the \underline{Top} dataset, (again, see figure~\ref{fig:multiplicity}),
is by far the best of any algorithm.
The flexibility of the clustering process allows it to separate jets that lie close together
while still gathering enough mass from jets in sparser areas to pass the mass cuts.

\subsection{Mass peak reconstruction with pileup}

Now the investigations of section~\ref{sec:without} are repeated with pileup in the data.
As mentioned in section \ref{sec:particle_data}, the number of pileup vertices is drawn from a passion distribution with mean \(50\).
All the same tagging and mass peak constructions are used, so that
the two datasets can be directly compared.
Pileup from charged tracks that originate away from the primary vertex is removed,
as a common pileup mitigation technique~\cite{pileup_mitigation2019}.
All parameters of the \spectral{} algorithm are left the same as before.

\begin{figure}[htp]
    \includegraphics[width=1.\textwidth]{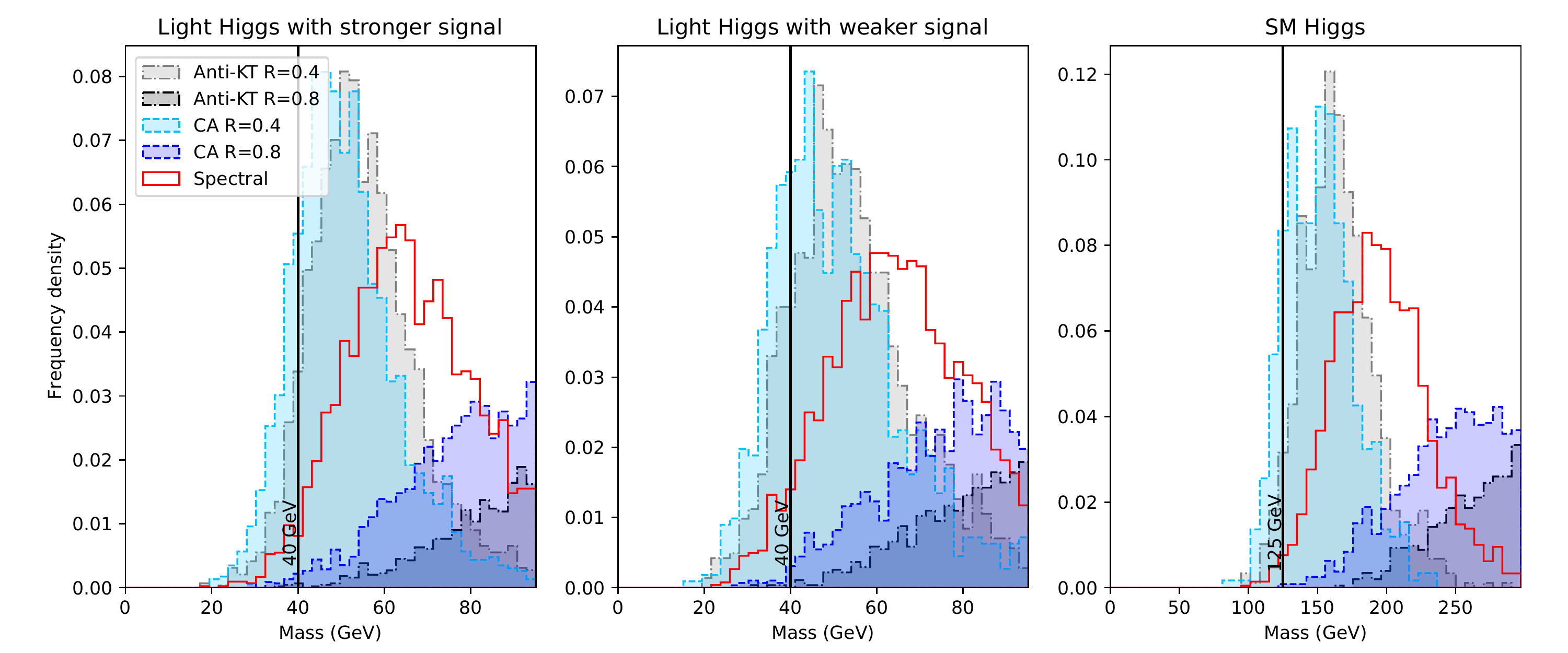}
    \caption{Three mass selections for the \underline{Light Higgs} dataset with pileup, as in figure~\ref{fig:best_correct_h_allocation}.
    }\label{fig:best_correct_h_allocation_pileup}
\end{figure}    

\begin{figure}[htp]
    \begin{center}
        \begin{minipage}[c]{0.46\textwidth}
        \includegraphics[width=1.\textwidth]{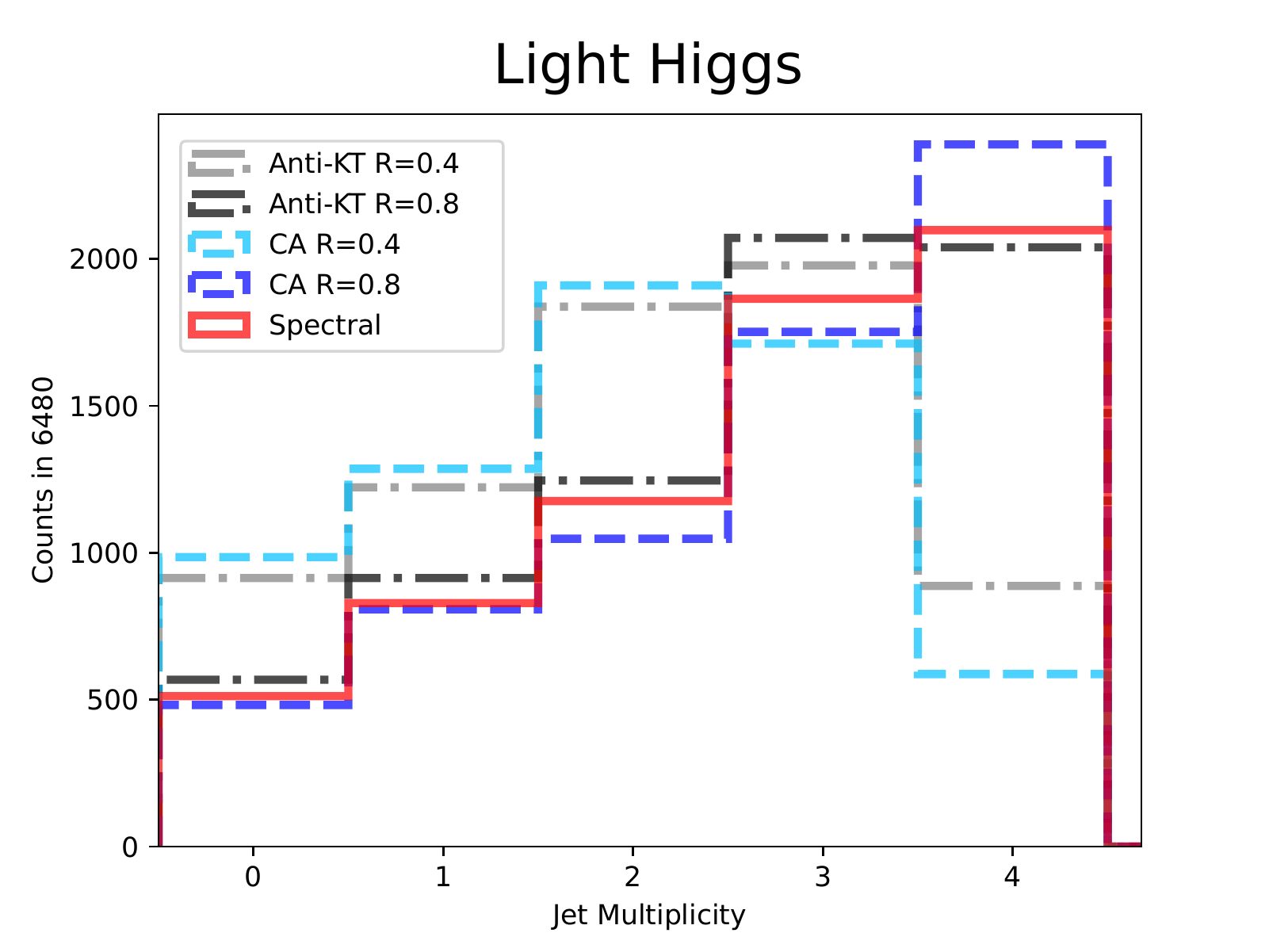}\\
        \includegraphics[width=1.\textwidth]{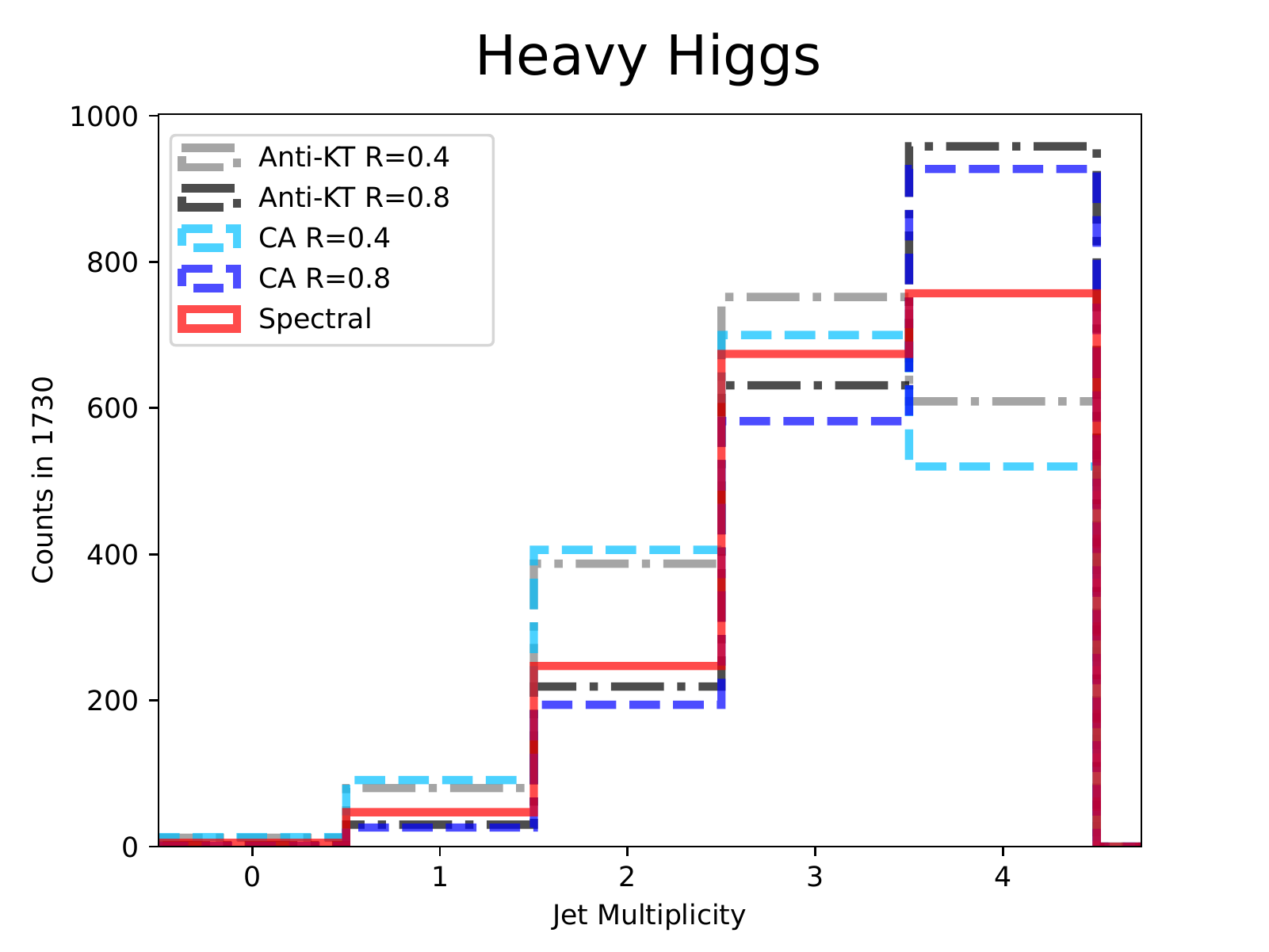}
        \end{minipage}\hfill
        \begin{minipage}[c]{0.46\textwidth}
        \includegraphics[width=1.\textwidth]{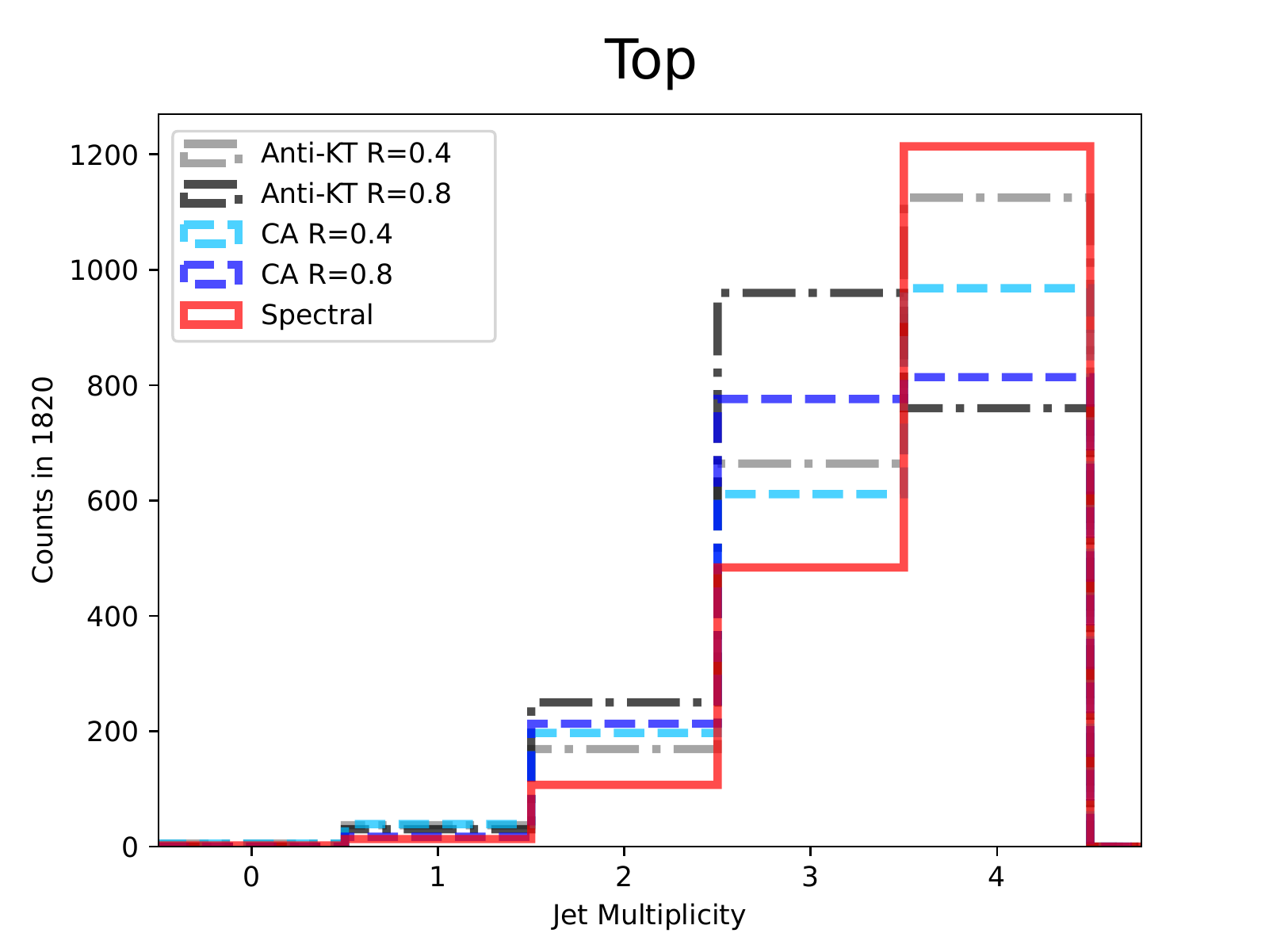}
        \caption{Jet multiplicities, with pileup, as in figure~\ref{fig:multiplicity}.
    }\label{fig:multiplicity_pileup}
        \end{minipage}
    \end{center}
\end{figure}

In figure~\ref{fig:best_correct_h_allocation_pileup} the \underline{Light Higgs} mass peaks are shown.
All jets have suffered, but it can be seen that \antikt{}/CA with \(R_{k_T}=0.4\) now produces
narrower and better positioned peaks than those of \(R_{k_T}=0.8\).
In the presence of pileup, jets with a wider joining distance become easily contaminated, and so gain too much mass.
Spectral clustering is not quite insensitive to this effect, but it has not entirely lost its shape,
as \antikt{}/CA with \(R_{k_T}=0.8\) has.
A broader peak has been created at somewhat too high a mass. 
Jet grooming would be needed to produce an acceptable peak here.
In the multiplicities plotted in figure~\ref{fig:multiplicity_pileup} it can be
seen that \spectral{} still has a multiplicity comparable to \antikt{}/CA with \(R_{k_T}=0.4\),
now substantially outperforming \(R_{k_T}=0.8\) which is suffering from merging jets.

\begin{figure}[htp]
    \includegraphics[width=1.\textwidth]{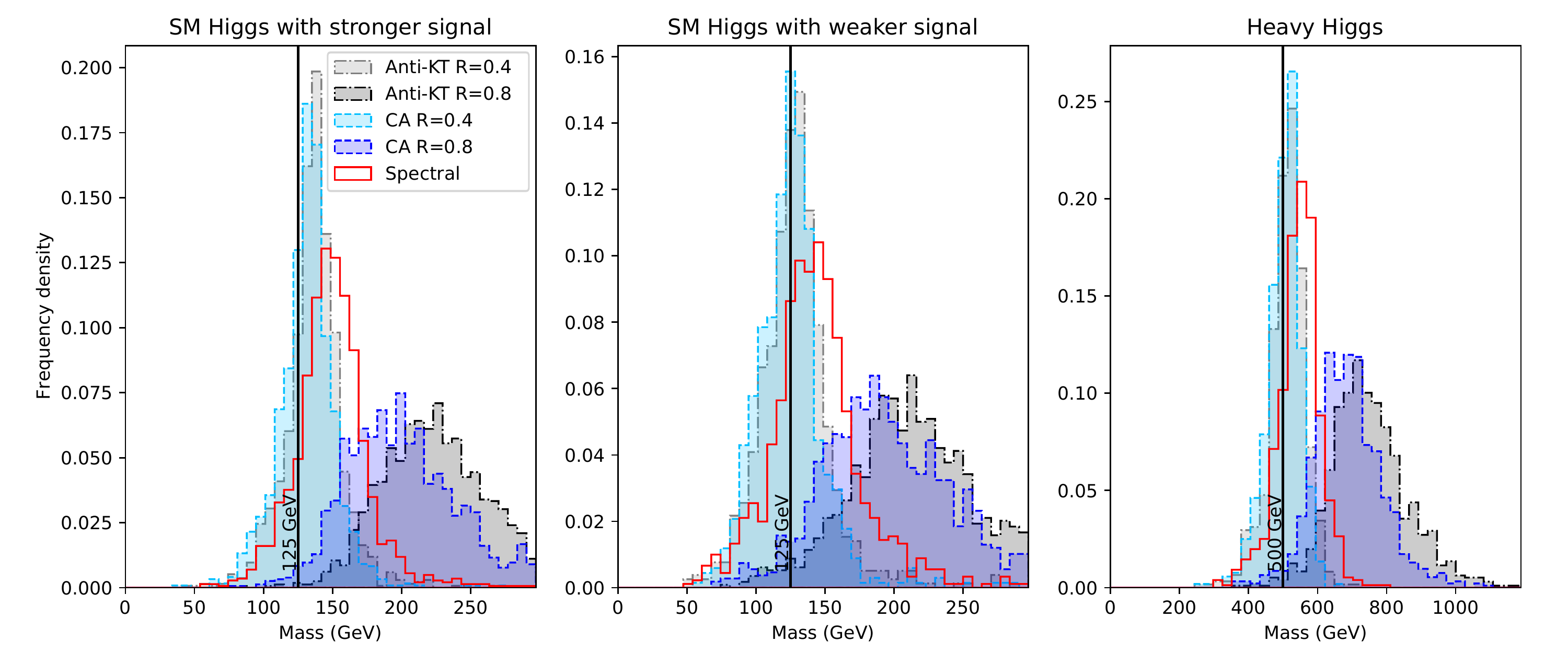}
    \caption{Three mass selections for the \underline{Heavy Higgs} dataset with pileup, as in figure~\ref{fig:heavy_correct_mass_peaks}.
    }\label{fig:heavy_correct_mass_peaks_pileup}
\end{figure}

Moving on to the \underline{Heavy Higgs} case, the mass peaks are presented in figure~\ref{fig:heavy_correct_mass_peaks_pileup}.
By comparison to figure~\ref{fig:heavy_correct_mass_peaks}, where \spectral{} clustering closely mimicked the \(R_{k_T}=0.8\)
cases, it can be seen that \spectral{} clustering has mostly avoided over-clustering.
Although it has taken on a little extra mass, the
peaks have not moved so far from where they were located without  pileup.
Any peak is indeed only a little broader than it was without pileup.
In figure~\ref{fig:multiplicity_pileup} it can be seen that the multiplicity for spectral clustering is not as good as for
the \(R_{k_T}=0.8\) methods, but it still is a fair performer and a good improvement on \(R_{k_T}=0.4\).

In summary, although it is clear that \antikt{} or CA with \(R_{k_T}=0.4\) is preferable overall in these cases,
\spectral{} clustering has done a respectable job of adapting, substantially outperforming the 0.8 cases. 

\begin{figure}[htp]
    \includegraphics[width=1.\textwidth]{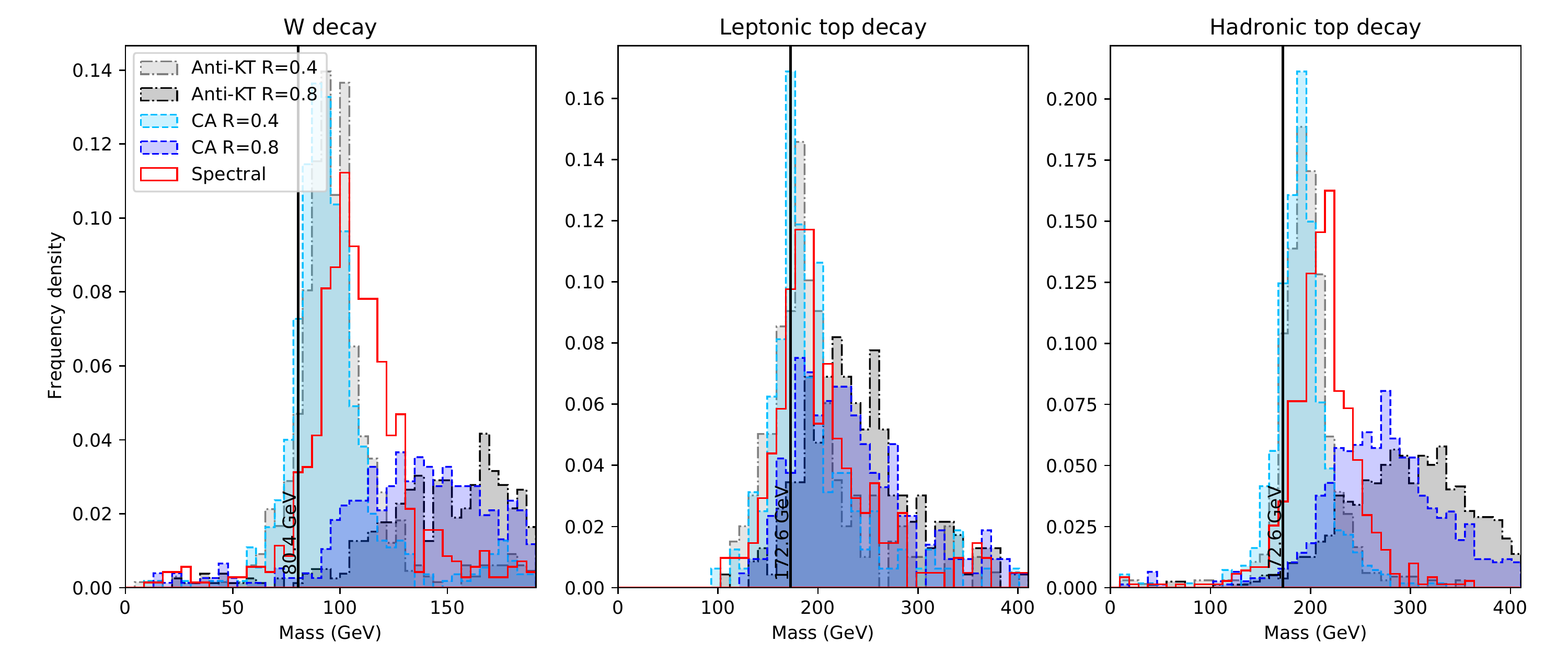}
    \caption{Three mass selections for the \underline{Top} dataset with pileup, as in figure~\ref{fig:top_correct_mass_peaks}.
    }\label{fig:top_correct_mass_peaks_pileup}
\end{figure}

Finally, the mass peaks for the \underline{Top} dataset with pileup are presented in figure~\ref{fig:top_correct_mass_peaks_pileup}.
This shows dramatically why \antikt{} with \(R_{k_T}=0.4\) is preferred to \antikt{} \(R_{k_T}=0.8\) for clustering semileptonic top decays.
Here, the mass peaks created by the \spectral{} algorithm are very nearly undamaged compared to the same without pileup in
figure~\ref{fig:top_correct_mass_peaks}.
There is a shift towards higher masses, but not particularly substantial compared to \antikt{}/CA with \(R_{k_T}=0.8\) and they remain quite sharp.
Looking back to the multiplicities with pileup in figure~\ref{fig:multiplicity_pileup} there is
excellent multiplicity from  \spectral{} algorithm.
Again, it exceeds all other choices at successfully isolating the jets, unhindered by the varying widths.

\subsection{Run Time}

Given the requirement for an eigenvalue calculation, an \(\mathcal{O}({n^3})\) operation,
it is clear that the spectral clustering algorithm will have longer runtimes than the \genkt{}
algorithm, which boasts \(\mathcal{O}({n\log(n)})\)~\cite{CACCIARI200657}.
The initial steps of the spectral algorithm require similar calculations to
\genkt{}, so it would be expected to have the same runtime.
The implementation used in this work actually neglects the
improvements that took \genkt{} from \(\mathcal{O}({n^3})\) to \(\mathcal{O}({n\log(n)})\),
so the time complexity should be at least \(\mathcal{O}({n^3})\).
However, an eigenvector calculation is typically \(\mathcal{O}({n^3})\), and this
may be repeated up to \(n\) times.
So with a na\"ive implementation, one would expect the spectral algorithm
to require \(\mathcal{O}({n^4})\).

\begin{figure}
    \begin{center}
        \begin{minipage}[c]{0.65\textwidth}
    \includegraphics[width=1.\textwidth]{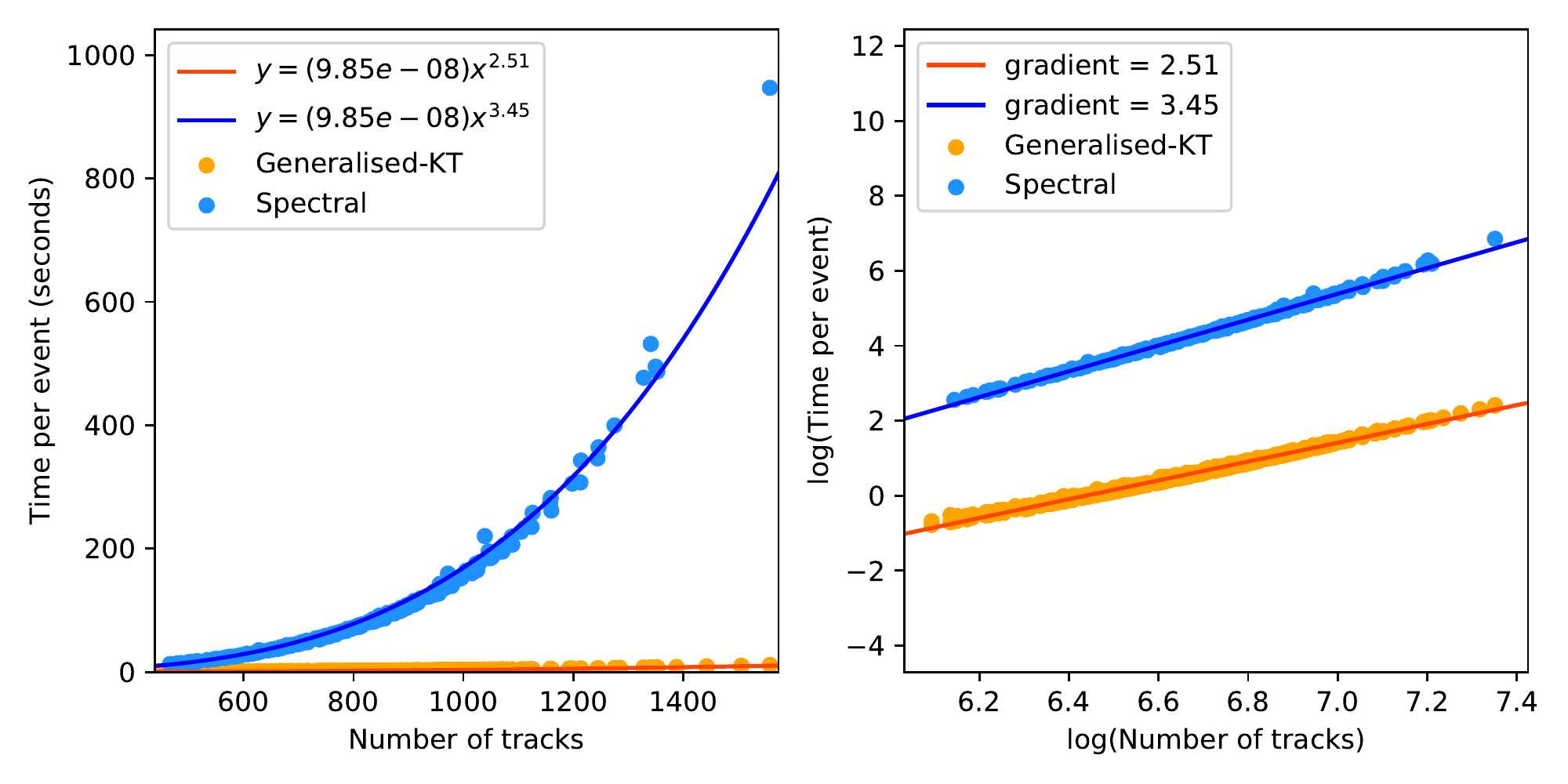}
        \end{minipage}\hfill
        \begin{minipage}[c]{0.35\textwidth}
    \caption{The run time of \spectral{} clustering compared to a na\"ive implementation of
        \genkt{} (without the performance refinements in~\cite{CACCIARI200657}),
        on datasets of varying size.
        Simple fits are shown for each dataset, in both linear and logarithmic scale.
        This shows that \spectral{} clustering runs in just over \(\mathcal{O}({n^3})\).
        }\label{fig:run_time}
        \end{minipage}
\end{center}
\end{figure}

This reasoning makes the results in figure~\ref{fig:run_time} a little surprising.
Herein, it is seen that \spectral{} clustering in fact runs in a little over \(\mathcal{O}({n^3})\),
not \(\mathcal{O}({n^4})\).
No particular optimisations were used to achieve this,
the implementation of \spectral{} clustering is a basic python script of the
algorithm set out in \autoref{sec:spectralmethodalgo}.
Specifically, no effort was made to take advantage of the sparse Laplacian matrix
when performing the eigenvector calculation.
Indeed, if anything, the implementation contains more branches than required,
because it was designed to facilitate investigating variations, such
as those shown in figure~\ref{fig:scan_spectral}.
The eigenvector calculation was performed by the \lstinline{scipy}'s~\cite{2020SciPy-NMeth}, 
function \lstinline{scipy.linalg.eigh}.

Nonetheless, further improvements to the run time would be needed to render this a practical algorithm.
Yet, this is outside the scope of this study.

    \FloatBarrier
    \section{Conclusions}

Spectral clustering is a popular machine learning algorithm, wherein complex datasets are transformed to clarify groupings in a new space.
In performing this transformation, it makes use of the spectrum (eigenvalues/eigenvectors) of the Laplacian matrix, which is constructed from localised information.
At no point in the process are large matrices of learnt parameters, common to deep learning methods, needed.
As such, spectral clustering is a transparent, simple to implement, algorithm using standard linear algebra methods. 
Owing to these features,  we have found it to also be 
a promising new method to apply to jet formation in high energy particle physics events.

For a start, it satisfies the need for IR safety and creates jets with the expected kinematics, as dictated by QCD dynamics. Furthermore, 
while it has many parameters, they do not appear to be as finely tuned as those of more standard tools, such us sequential (or iterative) \genkt{} algorithms.
This can be seen in both parameter scan stability and its adaptability to various datasets,
each capturing physics signals embedding heavy objects decaying into lighter ones in very different patterns, all yielding complicated hadronic signatures at the LHC.

The adaptability between datasets is remarkable as a \spectral{} clustering parameter choice tuned on a light Higgs boson cascade
gave excellent performance on both a heavy Higgs boson cascade and that of top-antitop pairs decaying semi-leptonically.
In the case of the \underline{Light Higgs} dataset, \spectral{} clustering gave the correct mass peak positions, the narrowest resonant distributions and a jet multiplicity mapping well the partonic one. This would not be surprising as it was tuned for that dataset in the first place.
In the case of the \underline{Heavy Higgs} dataset only \antikt{} with \(\ktstoppingdeltar{} = 0.8\) 
and the \spectral{} algorithm gave correct mass peaks but  \spectral{} clustering offers considerably better multiplicity rates.
This demonstrates that its performance is not dependent on fine tuning its parameters and hence that the algorithm is adaptable to the same final state with different masses involved.
Finally, \spectral{} clustering was applied to a \underline{Top} dataset with a different final state and 
for which the ideal jet radius differed, i.e., semileptonic decays of top-antitop pairs.
Its equivalent parameter \(\sigma_v\) was not allowed to vary to account for this, instead it was applied again with no parameter changes.
The algorithm again proved to be adaptable and modified its behaviour to follow that of anti-$k_T$ with \(\ktstoppingdeltar{} = 0.4\), the standard choice for this kind of analyses.

Pileup was seen to drastically alter the mass peaks formed by \antikt{} algorithms:
this is a well-known challenge. 
What was interesting to see was that the addition of pileup did not alter the performance of the \spectral{}  algorithm
as drastically as it did to those of the \antikt{} and CA algorithms that were favourable without it.
Again, this evidences a flexibility, and relative insensitivity, to the specifics of the clusters needed.

In short, spectral clustering is a novel and promising approach to jet formation, whose initial development already demonstrates flexibility and excellent performance for numerical analyses at the forefront of collider physics, open to further improvements, including a faster
implementation, that will be the subject of future publications.

    \FloatBarrier

    \section{Acknowledgements}
We thank A. Chakraborty, 
J. Chaplais, 
S. Jain and 
E. Olaiya for insightful discussions. HAD-H thanks
G.P. Salam for useful advice. 
HAD-H, BF, SM and CHS-T are supported in part through the NExT Institute.
SM is also supported by the STFC Consolidated
Grant No. ST/L000296/1. BF is funded by the DISCnet \& SEPnet scholarship schemes.
We finally 
acknowledge the use of the IRIDIS High Performance Computing Facility, and associated
support services, at the University of Southampton, in the completion of this work.

    \appendix
    \section{Stopping condition}\label{app:stopping_conditions}

\begin{figure}[!t]
        \center
        \includegraphics[width=\textwidth]{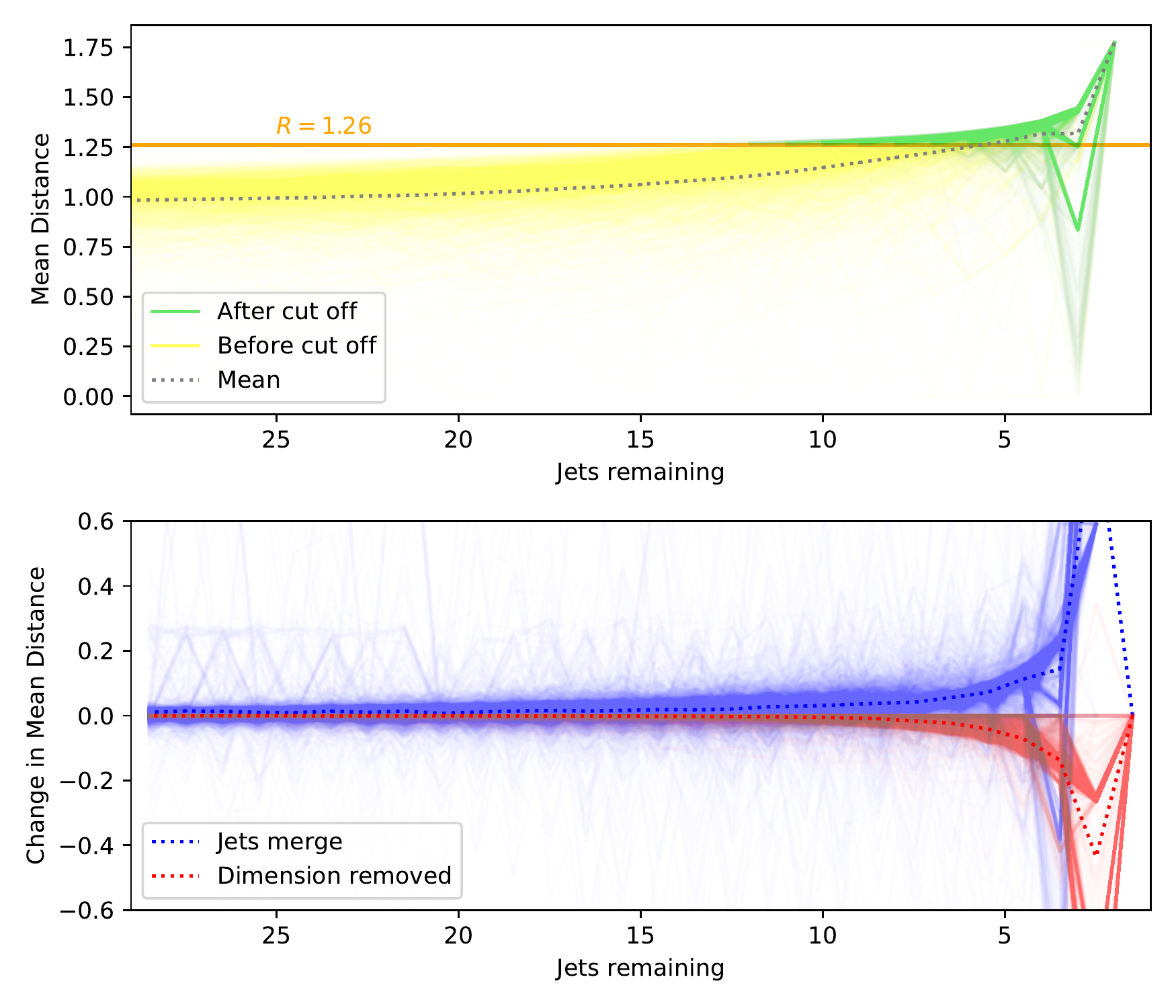}
        \caption{
        In the upper panel, the mean distance between pseudojets for \(2000\) events is plotted against the number of pseudojets remaining.
        Each line is shown in yellow until its value first exceeds \(R=1.26\),
        the stopping condition, after which the line becomes green.
    A dotted line shows the average mean distance across all \(2000\) events.
        In the lower panel, the factors that alter the mean distance are plotted.
        Again, each of the \(2000\) events is represented as a single line, and the 
        average is given as a dotted line.
        In blue, change of mean distance due to merging pseudojets is shown.
        In red, change of mean distance due to a reduction in the number of dimensions in the embedding space
        is shown.
        }\label{fig:mean_distance_change}
    \end{figure}

To offer some evidence for the assertions made in section~\ref{sec:stopping_condintion},
the behaviour of the mean distance during clustering is shown in figure~\ref{fig:mean_distance_change}.

Clustering is performed on the dataset described in section~\ref{sec:particle_data} called \underbar{Light Higgs}.
The parameters used for the \spectral{} algorithm are the ones given at the end of section~\ref{sec:spectralmethodparam}.
First, the upper panel of figure~\ref{fig:mean_distance_change} shows the mean distance between pseudojets for \(2000\) events,
plotted against the number of pseudojets remaining.
Each line is shown in yellow until its value first exceeds \(R=1.26\),
the stopping condition, after which the line becomes green.
When finding jets with \spectral{} clustering,
the algorithm would normally be stopped at the end of the yellow section,
as the stopping condition has been reached,
the green section is shown here to illustrate what happens beyond this point.
It can be seen that the transition from yellow to green happens with approximately \(3\) to \(13\) pseudojets remaining.
This supports the assertion that a mean distance stopping condition will not force the same number of jets in each event.
It can also be seen that the mean distance does rise smoothly for most of the clustering sequence,
becoming erratic only when less than \(5\) pseudojets remain.

Second, in the lower panel, the factors that alter the mean distance are plotted.
Again, each of the \(2000\) events is represented as a single solid line.
In blue, change of mean distance due to merging pseudojets is shown.
Normally merging two pseudojets causes the mean distance to rise,
as the embedding space is becoming sparser, however,
there are some configurations in which this does not hold.
Occasionally, two points that merge will lower the mean distance,
and the blue line will dip below zero.
It can be seen from the plot that such configurations are less common than 
those that increase mean distance.

The second panel also shows change of mean distance due to a reduction in the number of dimensions in the embedding space in red.
This universally decreases mean distance, the red lines remain below or at zero.
Not every step of the algorithm will reduce the number of dimensions,
and so the red line for an event is frequently zero.

It can be seen that these two factors balance each other to 
produce a steady trend in mean distance.

There is a third possibility, very rarely the number of dimensions in the 
embedding space will increase.
This is not pictured, as it is not possible to visually distinguish the line from \(y=0\)
and it would clutter the plot.

\bibliography{spectraljet}   

\providecommand{\href}[2]{#2}\begingroup\raggedright\begin{thebibliography}{10}

\bibitem{Ellis:1993tq}
S.D.~Ellis and D.E.~Soper, \emph{{Successive combination jet algorithm for
  hadron collisions}},
  \href{https://doi.org/10.1103/PhysRevD.48.3160}{\emph{Phys. Rev.} {\bfseries
  D48} (1993) 3160} [\href{https://arxiv.org/abs/hep-ph/9305266}{{\ttfamily
  hep-ph/9305266}}].

\bibitem{Dokshitzer:1997in}
Y.L.~Dokshitzer, G.D.~Leder, S.~Moretti and B.R.~Webber, \emph{{Better jet
  clustering algorithms}},
  \href{https://doi.org/10.1088/1126-6708/1997/08/001}{\emph{JHEP} {\bfseries
  08} (1997) 001} [\href{https://arxiv.org/abs/hep-ph/9707323}{{\ttfamily
  hep-ph/9707323}}].

\bibitem{Wobisch:1998wt}
M.~Wobisch and T.~Wengler, \emph{{Hadronization corrections to jet
  cross-sections in deep inelastic scattering}},  in \emph{{Monte Carlo
  generators for HERA physics. Proceedings, Workshop, Hamburg, Germany,
  1998-1999}}, pp.~270--279, 1998
  [\href{https://arxiv.org/abs/hep-ph/9907280}{{\ttfamily hep-ph/9907280}}].

\bibitem{Cacciari:2008gp}
M.~Cacciari, G.P.~Salam and G.~Soyez, \emph{{The anti-$k_t$ jet clustering
  algorithm}}, \href{https://doi.org/10.1088/1126-6708/2008/04/063}{\emph{JHEP}
  {\bfseries 04} (2008) 063} [\href{https://arxiv.org/abs/0802.1189}{{\ttfamily
  0802.1189}}].

\bibitem{Catani:1993hr}
S.~Catani, Y.L.~Dokshitzer, M.H.~Seymour and B.R.~Webber, \emph{{Longitudinally
  invariant $K_t$ clustering algorithms for hadron hadron collisions}},
  \href{https://doi.org/10.1016/0550-3213(93)90166-M}{\emph{Nucl. Phys. B}
  {\bfseries 406} (1993) 187}.

\bibitem{Moretti:1998qx}
S.~Moretti, L.~Lonnblad and T.~Sjostrand, \emph{{New and old jet clustering
  algorithms for electron - positron events}},
  \href{https://doi.org/10.1088/1126-6708/1998/08/001}{\emph{JHEP} {\bfseries
  08} (1998) 001} [\href{https://arxiv.org/abs/hep-ph/9804296}{{\ttfamily
  hep-ph/9804296}}].

\bibitem{Sterman:1977wj}
G.F.~Sterman and S.~Weinberg, \emph{{Jets from Quantum Chromodynamics}},
  \href{https://doi.org/10.1103/PhysRevLett.39.1436}{\emph{Phys. Rev. Lett.}
  {\bfseries 39} (1977) 1436}.

\bibitem{Bethke:1991wk}
S.~Bethke, Z.~Kunszt, D.E.~Soper and W.J.~Stirling, \emph{{New jet cluster
  algorithms: Next-to-leading order QCD and hadronization corrections}},
  \href{https://doi.org/10.1016/0550-3213(92)90289-N}{\emph{Nucl. Phys.}
  {\bfseries B370} (1992) 310}.

\bibitem{Catani:1991hj}
S.~Catani, Y.L.~Dokshitzer, M.~Olsson, G.~Turnock and B.R.~Webber, \emph{{New
  clustering algorithm for multi - jet cross-sections in $e+$ $e-$
  annihilation}},
  \href{https://doi.org/10.1016/0370-2693(91)90196-W}{\emph{Phys. Lett.}
  {\bfseries B269} (1991) 432}.

\bibitem{Cacciari:2011ma}
M.~Cacciari, G.P.~Salam and G.~Soyez, \emph{{FastJet User Manual}},
  \href{https://doi.org/10.1140/epjc/s10052-012-1896-2}{\emph{Eur. Phys. J. C}
  {\bfseries 72} (2012) 1896}
  [\href{https://arxiv.org/abs/1111.6097}{{\ttfamily 1111.6097}}].

\bibitem{Belkin:2003_unfound4}
M.~Belkin and P.~Niyogi, \emph{Laplacian eigenmaps for dimensionality reduction
  and data representation},
  \href{https://doi.org/10.1162/089976603321780317}{\emph{Neural Comput.}
  {\bfseries 15} (2003) 1373–1396}.

\bibitem{Shi:1997_unfound595}
J.~Shi and J.~Malik, \emph{Normalized cuts and image segmentation},  in
  \emph{Proceedings of the 1997 Conference on Computer Vision and Pattern
  Recognition (CVPR '97)}, CVPR '97, (USA), p.~731, IEEE Computer Society,
  1997.

\bibitem{Ng:2001_unfound543}
A.Y.~Ng, M.I.~Jordan and Y.~Weiss, \emph{On spectral clustering: Analysis and
  an algorithm},  in \emph{Proceedings of the 14th International Conference on
  Neural Information Processing Systems: Natural and Synthetic}, NIPS'01,
  (Cambridge, MA, USA), p.~849–856, MIT Press, 2001.

\bibitem{Hadjighasem:2016_unfound447}
A.~Hadjighasem, D.~Karrasch, H.~Teramoto and G.~Haller,
  \emph{Spectral-clustering approach to lagrangian vortex detection},
  \href{https://doi.org/10.1103/PhysRevE.93.063107}{\emph{Phys. Rev. E}
  {\bfseries 93} (2016) 063107}
  [\href{https://arxiv.org/abs/1506.02258}{{\ttfamily 1506.02258}}].

\bibitem{HaoLi:2005_unfound114}
{Hao Li}, G.W.~{Rosenwald}, J.~{Jung} and {Chen-ching Liu}, \emph{Strategic
  power infrastructure defense},
  \href{https://doi.org/10.1109/JPROC.2005.847260}{\emph{Proceedings of the
  IEEE} {\bfseries 93} (2005) 918}.

\bibitem{RJSanchezGarcia:2014_unfound420}
R.J.~{S\'{a}nchez-Garc\'{i}a}, M.~{Fennelly}, S.~{Norris}, N.~{Wright},
  G.~{Niblo}, J.~{Brodzki} et~al., \emph{Hierarchical spectral clustering of
  power grids}, {\emph{IEEE Transactions on Power Systems} {\bfseries 29}
  (2014) 2229}.

\bibitem{UlrikevonLuxburg:2007_unfound52}
U.~von Luxburg, \emph{A tutorial on spectral clustering},  2007.

\bibitem{Leeuwen:1990_unfound0}
J.V.~Leeuwen, Warwick, A.R.~Meyer and M.~Nival, \emph{Handbook of Theoretical
  Computer Science: Algorithms and Complexity}, MIT Press, Cambridge, MA, USA
  (1990).

\bibitem{JamesRLee:2014_unfound736}
J.R.~Lee, S.O.~Gharan and L.~Trevisan, \emph{Multiway spectral partitioning and
  higher-order cheeger inequalities},
  \href{https://doi.org/10.1145/2665063}{\emph{J. ACM} {\bfseries 61} (2014) }.

\bibitem{Ju:2020tbo}
X.~Ju and B.~Nachman, \emph{{Supervised Jet Clustering with Graph Neural
  Networks for Lorentz Boosted Bosons}},
  \href{https://doi.org/10.1103/PhysRevD.102.075014}{\emph{Phys. Rev. D}
  {\bfseries 102} (2020) 075014}
  [\href{https://arxiv.org/abs/2008.06064}{{\ttfamily 2008.06064}}].

\bibitem{Chakraborty:2020vwj}
A.~Chakraborty, S.~Dasmahapatra, H.~Day-Hall, B.~Ford, S.~Jain, S.~Moretti
  et~al., \emph{{Revisiting Jet Clustering Algorithms for New Higgs Boson
  Searches in Hadronic Final States}},
  \href{https://arxiv.org/abs/2008.02499}{{\ttfamily 2008.02499}}.

\bibitem{Moretti:1994ds}
S.~Moretti and W.J.~Stirling, \emph{{Contributions of below threshold decays to
  MSSM Higgs branching ratios}},
  \href{https://doi.org/10.1016/0370-2693(95)00088-3,
  10.1016/0370-2693(95)01477-2}{\emph{Phys. Lett.} {\bfseries B347} (1995) 291}
  [\href{https://arxiv.org/abs/hep-ph/9412209}{{\ttfamily hep-ph/9412209}}].

\bibitem{Djouadi:1995gv}
A.~Djouadi, J.~Kalinowski and P.M.~Zerwas, \emph{{Two and three-body decay
  modes of SUSY Higgs particles}},
  \href{https://doi.org/10.1007/s002880050121}{\emph{Z. Phys.} {\bfseries C70}
  (1996) 435} [\href{https://arxiv.org/abs/hep-ph/9511342}{{\ttfamily
  hep-ph/9511342}}].

\bibitem{Alwall:2011uj}
J.~Alwall, M.~Herquet, F.~Maltoni, O.~Mattelaer and T.~Stelzer, \emph{Madgraph
  5: going beyond},
  \href{https://doi.org/10.1007/jhep06(2011)128}{\emph{Journal of High Energy
  Physics} {\bfseries 2011} (2011) 128}
  [\href{https://arxiv.org/abs/1106.0522}{{\ttfamily 1106.0522}}].

\bibitem{Sjostrand:2014zea}
T.~Sjostrand, S.~Ask, J.R.~Christiansen, R.~Corke, N.~Desai, P.~Ilten et~al.,
  \emph{An introduction to {PYTHIA} 8.2},
  \href{https://arxiv.org/abs/1410.3012}{{\ttfamily 1410.3012}}.

\bibitem{pileup_mitigation2019}
G.~Soyez, \emph{Pileup mitigation at the lhc: A theorist’s view},
  \href{https://doi.org/10.1016/j.physrep.2019.01.007}{\emph{Physics Reports}
  {\bfseries 803} (2019) 1–158}.

\bibitem{Sirunyan:2018fpa}
A.~Sirunyan, A.~Tumasyan, W.~Adam, F.~Ambrogi, E.~Asilar, T.~Bergauer et~al.,
  \emph{Performance of the cms muon detector and muon reconstruction with
  proton-proton collisions at $\sqrt{s}=13$ tev},
  \href{https://doi.org/10.1088/1748-0221/13/06/p06015}{\emph{Journal of
  Instrumentation} {\bfseries 13} (2018) P06015–P06015}
  [\href{https://arxiv.org/abs/1804.04528}{{\ttfamily 1804.04528}}].

\bibitem{Sirunyan:2019rfa}
A.M.~Sirunyan, A.~Tumasyan, W.~Adam, F.~Ambrogi, T.~Bergauer, J.~Brandstetter
  et~al., \emph{Measurement of the jet mass distribution and top quark mass in
  hadronic decays of boosted top quarks in $pp$ collisions at $\sqrt{s}=13$
  tev}, \href{https://doi.org/10.1103/physrevlett.124.202001}{\emph{Physical
  Review Letters} {\bfseries 124} (2020) }
  [\href{https://arxiv.org/abs/1911.03800}{{\ttfamily 1911.03800}}].

\bibitem{Lin:1991zzm}
J.~{Lin}, \emph{Divergence measures based on the shannon entropy},
  \href{https://doi.org/10.1109/18.61115}{\emph{IEEE Transactions on
  Information Theory} {\bfseries 37} (1991) 145}.

\bibitem{Altarelli:1989hv}
CERN, \emph{{$Z$ Physics at LEP1: CERN, Geneva, Switzerland 20 - 21 Feb, 8 - 9
  May and 4 - 5 Sep 1989.}}, (Geneva), CERN, 1989.
\newblock 10.5170/CERN-1989-008-V-1.

\bibitem{CACCIARI200657}
M.~Cacciari and G.P.~Salam, \emph{Dispelling the $n^3$ myth for the $k_t$
  jet-finder},
  \href{https://doi.org/https://doi.org/10.1016/j.physletb.2006.08.037}{\emph{Physics
  Letters B} {\bfseries 641} (2006) 57}.

\bibitem{2020SciPy-NMeth}
P.~Virtanen, R.~Gommers, T.E.~Oliphant, M.~Haberland, T.~Reddy, D.~Cournapeau
  et~al., \emph{{{SciPy} 1.0: Fundamental Algorithms for Scientific Computing
  in Python}}, \href{https://doi.org/10.1038/s41592-019-0686-2}{\emph{Nature
  Methods} {\bfseries 17} (2020) 261}.

\end{thebibliography}\endgroup
\end{document}